\global\def\draftcontrol{0}

   \def\versionno{ bbb}
\catcode`\@=11

\expandafter\ifx\csname draftcontrol\endcsname\relax\global\def\draftcontrol{0}
\fi

{\count255=\time\divide\count255 by 60
\xdef\hourmin{\number\count255}
\multiply\count255 by-60\advance\count255 by\time
\xdef\hourmin{\hourmin:\ifnum\count255<10 0\fi\the\count255}}
\def\draftdate{\number\month/\number\day/\number\year\ \ \ \hourmin }

\newcommand\makepapertitle{\par
  \begingroup
    \renewcommand\thefootnote{\@fnsymbol\c@footnote}%
    \def\@makefnmark{\rlap{\@textsuperscript{\normalfont\@thefnmark}}}%
    \long\def\@makefntext##1{\parindent 1em\noindent
            \hb@xt@1.8em{%
                \hss\@textsuperscript{\normalfont\@thefnmark}}##1}%
     \newpage
     \global\@topnum\z@   % Prevents figures from going at top of page.
     \@makepapertitle
     \thispagestyle{empty}\@thanks
  \endgroup
  \setcounter{footnote}{0}%
  \global\let\thanks\relax
  \global\let\makepapertitle\relax
  \global\let\@makepapertitle\relax
  \global\let\@thanks\@empty
  \global\let\@author\@empty
  \global\let\@date\@empty
  \global\let\@title\@empty
  \global\let\title\relax
  \global\let\author\relax
  \global\let\date\relax
  \global\let\and\relax
  \def\version{\let\version\@version\@gobble}
}
\def\@makepapertitle{%
  \newpage
   \ifnum\draftcontrol=1 {}
   \version\versionno
   \vskip 3em%
   \else
   \hfill\hbox to 3cm {\parbox{4cm}{\@pubnum}\hss}%
   \vskip 3em%
   \fi
   \begin{center}%
   \let \footnote \thanks
     {\LARGE {\@title}}%
     \vskip 1.5em%
     {\normalsize%\large
       \lineskip .5em%
       \begin{tabular}[t]{c}%
         \@author
       \end{tabular}\par}%
     \vskip 1.5em%
     {\@bstract}%
     \end{center}%
     \vskip 1.5em
     \@date%
   \par
}

\gdef\@pubnum{}
\def\pubnum#1{%
  \gdef\@pubnum{#1}}

\gdef\@bstract{}
\def\Abstract#1{%
  \gdef\@bstract{%
   \parbox{\textwidth-0pc}{%
   \centerline{\bf Abstract}\penalty1000%
\kern.2cm%
\noindent%\abstractfont \baselineskip=12pt
\renewcommand\baselinestretch{1.0}%
{#1}}}
}

\def\ps@paper{\let\@mkboth\@gobbletwo%
     \ifnum\draftcontrol=1
    \def\@oddfoot{\hbox to \textwidth{\tiny \versionno \hfil\tiny\draftdate}%
    \hskip -\textwidth \hbox to \textwidth{\hfil\rm\thepage\hfil}}%
     \else\def\@oddfoot{\hbox to \textwidth{\hfil\rm\thepage\hfil}}
     \fi
     \let\@evenfoot\@oddfoot
}
\def\body{\clearpage
          \pagestyle{paper}
    }
\def\@version#1{\ifnum\draftcontrol=1
\typeout{}\typeout{#1}\typeout{}
\vskip3mm\centerline{\hbox{\fbox{\normalsize{\tt DRAFT -- #1 -- }
                   {\draftdate}}}}\vskip3mm
\fi}
\let\version\@version
\long\def\eqlabel#1{\ifnum\draftcontrol=1
                    \tag@false  % there are some problems with multline without this
                    \tag*{(\theequation) \hbox to -0.2cm{\hspace{0cm}\small{#1}\hss}}
                    \refstepcounter{equation}
                    \edef\@currentlabel{\theequation}
                    \ltx@label{#1}          % use old LaTeX \label instead of new definition
                    \else
                    \label{#1}
                    \fi
                    }
\let\st@bibitem\@bibitem
\let\st@lbibitem\@lbibitem
\ifnum\draftcontrol=1
  \def\@bibitem#1{%
    \st@bibitem{#1}\a@@label{#1}\ignorespaces}
  \def\@lbibitem[#1]#2{%
    \st@lbibitem[#1]{#2}\a@@label{#2}\ignorespaces}
  \def\a@@label#1{%
    \gdef\a@lab{\smash{\normalfont\small#1}}
    \ifvmode
      \if@inlabel
        \global\setbox\@labels\hbox{%
          \llap{\a@lab\let\a@lab\relax
                \kern\@totalleftmargin\kern\marginparsep}%
          \box\@labels}%
      \fi
    \fi}
\fi
\documentclass[12pt,letterpaper]{article}
\usepackage{amsmath,amssymb,array,calc,epsfig,rotating,bm}
\usepackage[sort]{cite}
\usepackage{graphicx}
\usepackage{psfrag,verbatim}
\usepackage{xcolor}
\usepackage{hyperref}

\ifnum\draftcontrol=1
\fi
\renewcommand\baselinestretch{1.25}
\renewcommand\section{\@startsection {section}{1}{\z@}%
                                   {-3.5ex \@plus -1ex \@minus -.2ex}%
                                   {2.3ex \@plus.2ex}%
                                   {\normalfont\large\bfseries}}
\renewcommand\subsection{\@startsection{subsection}{2}{\z@}%
                                   {-3.25ex\@plus -1ex \@minus -.2ex}%
                                   {1.5ex \@plus .2ex}%
                                   {\normalfont\normalsize\bfseries}}
\renewcommand\subsubsection{\@startsection{subsubsection}{3}{\z@}%
                                   {-3.25ex\@plus -1ex \@minus -.2ex}%
                                   {1.5ex \@plus .2ex}%
                                   {\normalfont\normalsize\it}}
\renewcommand\paragraph{\@startsection{paragraph}{4}{\z@}%
                                   {-3.25ex\@plus -1ex \@minus -.2ex}%
                                   {1.5ex \@plus .2ex}%
                                   {\normalfont\normalsize\bf}}

\numberwithin{equation}{section}

\def\revise#1       {\raisebox{-0em}{\rule{3pt}{1em}}%
                     \marginpar{\raisebox{.5em}{\vrule width3pt\
                     \vrule width0pt height 0pt depth0.5em
                     \hbox to 0cm{\hspace{0cm}{%
                     \parbox[t]{4em}{\raggedright\footnotesize{#1}}}\hss}}}}

\newcommand\nxt[1]  {\\\fnxt#1}
\newcommand{\ie}{{\it i.e.,}\ }

\def\cala         {{\cal A}}

\def\cale         {{\cal E}}

\def\call         {{\cal L}}
\def\calm         {{\cal M}}
\def\caln         {{\cal N}}
\def\calo         {{\cal O}}

\def\cals         {{\cal S}}

\def\calv         {{\cal V}}

\def\reals        {{\mathbb R}}

\def\del          {\partial}

\def\Re           {{\rm Re\hskip0.1em}}
\def\Im           {{\rm Im\hskip0.1em}}

\def\sqr#1#2{{\vcenter{\vbox{\hrule height.#2pt
 \hbox{\vrule width.#2pt height#1pt \kern#1pt
 \vrule width.#2pt}\hrule height.#2pt}}}}

\newcommand{\ft}[2]{{\textstyle{\frac{#1}{#2}}}}

\newcommand{\kk}{\mathfrak{q}}
\newcommand{\ww}{\mathfrak{w}}

\def\aa1{\phi}
\def\cc1{\psi}

\def\Om{\Omega}

\catcode`\@=12

\begin{document}

%%%
%%%%%% text starts here
%%%%%%%%%

\title{\bf Instability of Baryonic Black Branes}

\date{February 9, 2025}
%\date\today

\author{
Alex Buchel\\[0.4cm]
\it Department of Physics and Astronomy\\ 
\it University of Western Ontario\\
\it London, Ontario N6A 5B7, Canada\\
\it Perimeter Institute for Theoretical Physics\\
\it Waterloo, Ontario N2J 2W9, Canada\\
}

\Abstract{Baryonic black branes describe the quantum critical phase of the
conformal conifold gauge theory at strong coupling. This phase extends
to zero temperature at a finite baryonic chemical potential,
represented by extremal black branes with $AdS_2\times R^3\times
T^{1,1}$ throat in asymptotic $AdS_5\times T^{1,1}$ geometry. We
demonstrate here that this phase is dynamically unstable below some
critical value of $T_c/\mu$: the instability is represented by a
diffusive mode in the hydrodynamic sound channel with a negative
diffusion coefficient.  We also identify a new (exotic) ordered phase
of the conifold gauge theory: this phase originates at the same
critical value of $T_c/\mu$, but extends to arbitrary high
temperatures, and is characterized by an expectation value of a
dimension-2 operator, $\calo_2\propto T^2$, in the limit $\frac \mu
T\to 0$.
}

\makepapertitle

\body

\version\versionno
\tableofcontents

\section{Introduction and summary}\label{intro}

In \cite{Herzog:2009gd} (HKPT) the authors discussed an interesting top-down model of the emergent holographic criticality
based on $\caln=1$ superconformal $SU(N)\times SU(N)$ Klebanov-Witten (KW) gauge theory  \cite{Klebanov:1998hh}. 
The gauge theory has $U(1)_R\times U(1)_B$ ($R$-charge and baryonic) global symmetry. There is a  phase
of the theory with a nonzero R-charge chemical potential $\mu_R\ne 0$ and a vanishing baryonic chemical
potential $\mu_B=0$, with a holographic dual represented by Reissner-Nordstrom black branes in
asymptotically $AdS_5$ space-time. While one can reach a quantum criticality, \ie the limit of
vanishing temperature with a finite entropy density, in this phase,
it was expected that the corresponding extremal black branes
would be unstable due to the presence of the charged under the $R$-symmetry matter
\footnote{Such instabilities were indeed identified in \cite{Buchel:2024phy}.}.
Instead, the authors of \cite{Herzog:2009gd} studied a phase of the KW gauge theory
with $\mu_R=0$ and $\mu_B\ne 0$ ---
the corresponding gravitational dual named ``baryonic black branes''. It was shown that quantum criticality
can be reached in this phase as well, and the authors conjectured 
that the absence of charged matter under $U(1)_B$ symmetry in type IIB supergravity
description\footnote{The gauge invariant operators with baryonic charge in KW gauge theory
have conformal dimension of order $N$, realized by wrapped D3-branes with charge-to-mass
ratio being too small to trigger the instability of the holographic superconductors \cite{Gubser:2009qm}. }
would make this phase perturbatively\footnote{Non-perturbative "Fermi
seasickness'' instability \cite{Hartnoll:2009ns} has been
known to authors of \cite{Herzog:2009gd} - I would like to thank Igor Klebanov
for stressing this point. We comment on this instability below.}
stable even in the limit $\frac {T}{\mu_B}\to 0$.

In this paper we establish that the HKPT quantum criticality is perturbatively
unstable after all.
The instability here parallels the recently identified instability of $\caln=4$
supersymmetric Yang-Mills (SYM) theory in a phase with a diagonal $U(1)_R$-symmetry
chemical potential \cite{Gladden:2024ssb}. Specifically, we find that
the HKPT gauge theory plasma is unstable for
\begin{equation}
\frac{T}{\mu_B}< \frac{T}{\mu_B}\bigg|_{crit}=0.2770(5)
\eqlabel{tmuc}
\end{equation}
to fluctuations of a combination of a neutral dimension-2 scalar
(in $\caln=4$ Betti vector multiple), a $U(1)_R$ gauge field, and a massive vector
of the $\caln=2$ consistent subtruncation of type IIB supergravity of warped
deformed conifold with fluxes \cite{Ceresole:1999zs,Cassani:2010na,Buchel:2014hja}.
The instability is a diffusive mode in the hydrodynamic sound channel with a dispersion
relation\footnote{We use notation $\ww\equiv \frac{\omega}{2\pi T}$ and $\kk\equiv \frac{|\vec q|}{2\pi T}$.} 
\begin{equation}
\ww=-i D\kk^2+\calo(\kk^2)\,,
\eqlabel{diff}
\end{equation} 
where the diffusion coefficient $D$ vanishes at critical temperature \eqref{tmuc}, and 
\begin{equation}
\begin{cases}
D>0\,,\qquad {\frac{T}{\mu_B}> \frac{T}{\mu_B}\bigg|_{crit}}\,,\\
D<0\,,\qquad {\frac{T}{\mu_B}< \frac{T}{\mu_B}\bigg|_{crit}}\,.
\end{cases}
\eqlabel{d}
\end{equation}

Much like in the case of $\caln=4$ supersymmetric Yang-Mills theory \cite{Buchel:2025cve,Buchel:2025tjq},
Klebanov-Witten gauge theory with a baryonic chemical potential has a spatially homogeneous exotic conformal ordered phase,
originating precisely at the critical temperature \eqref{tmuc}.
This phase extends to arbitrary high temperatures and is characterized by an expectation value
of a dimension-2 operator $\calo_2\ \propto T^2$ as $T\gg \mu_B$.
This phase, however, never dominates in the grand canonical ensemble as it has a higher
Gibbs free energy density than the corresponding HKPT baryonic black brane phase.
Exotic ordered phase of the $U(1)_B$ charged KW theory also carries $R$-symmetry
charge density --- the latter vanishes at critical temperature \eqref{tmuc}
as $\propto \sqrt{T-T_{crit}}$, exactly as the thermal expectation value of $\calo_2$ operator.
Further details of the phases of the conifold gauge theory at $\mu_B\ne 0$ and
$\mu_R=0$ are collected in section \ref{phase}.

As noted in \cite{Herzog:2009gd}, baryonic black branes suffer from the nonperturbative instability,
associated with the nucleation of the a space-time filling D-brane towards the AdS boundary. We revisit
this instability in section \ref{nonpert}, and identify its precise onset,
\begin{equation}
\frac{T}{\mu_B}<\frac{T}{\mu_B}\bigg|_{non-pert}=0.2789(9)\ \Longleftrightarrow\
\frac{T}{\mu_B}\bigg|_{non-pert}^{HKPT}=\frac{0.2789(9)}{\sqrt 2}=0.1972(7)\approx 0.2\,.
\eqlabel{tnonpert}
\end{equation}
The reported value \eqref{tnonpert} agrees with the one in \cite{Herzog:2009gd}:
there is a factor of $\sqrt{2}$ difference\footnote{I would like
to thank the referee for raising this possibility. See \eqref{sqrt2}.} between the normalization 
of the bulk gauge field (dual to $U(1)_B$ global symmetry) used here and in
\cite{Herzog:2009gd}.
While the brane nucleation instability is first triggered at a {\it higher} temperature 
than the perturbative instability \eqref{tmuc}, the brane nucleation rate is suppressed  
both in the large-$N$ 't Hooft limit, and by a spacial volume of a nucleated brane.

Section \ref{tech} collects the technical details. We begin with the review of the gravitational
dual effective action of the KW gauge theory with $U(1)_B\times U(1)_R$ global symmetries \cite{Cassani:2010na}. 
In section \eqref{hkpts} we review the baryonic black branes of \cite{Herzog:2009gd}.
In section \ref{hydro} we study the quasinormal modes of these baryonic black branes,
associated with the fluctuations of the $R$-symmetry charge density. We show that these fluctuations
become unstable below the critical temperature $\eqref{tmuc}$, leading to the clumping of the $R$-symmetry
charges along the translationary invariant horizon of HKPT black branes.
We continue the search for additional instabilities of baryonic black branes  in section \ref{onset}:
analyzing the fluctuation equations (derived in section \ref{hydro}) we predict a new homogeneous phase
of the conifold gauge theory plasma with a baryonic chemical potential, originating at
the critical temperature \eqref{tmuc}. This new phase is constructed in section \ref{normal}. 

It was argued in \cite{Buchel:2005nt} that the thermodynamic instabilities in neutral plasma
always imply the dynamical instabilities of the holographic dual black
branes (and of course
dynamical instabilities in the boundary gauge theory plasma itself!). Extension of this argument
to the charged plasma predicted the instabilities of the $\caln=4$ SYM charged plasma reported in 
\cite{Gladden:2024ssb}. For the KW gauge theory discussed here, we explicitly demonstrated the
dynamical instabilities, but we did not undertake the thermodynamic stability analysis
in the presence of $U(1)_B\times U(1)_R$ charge densities --- it would be interesting to fill this
gap\footnote{While we do not expect this to be the case in the model discussed,
dynamical instabilities in holographic models do not always imply the thermodynamic instabilities \cite{Buchel:2020jfs}.}.
Additionally, the ordered phase of the $\caln=4$ SYM plasma \cite{Buchel:2025cve} is
expected to be dynamically unstable \cite{Buchel:2025tjq,Gladden:2024ssb}. We do not know
if the corresponding ordered phase of the KW plasma is perturbatively stable.

We would like to conclude re-emphasizing the perils of near-extremal black branes.
It is natural to discuss gravitational quantum effects of the near-extremal horizons
in consistent theories of quantum gravity. Embedding near-extremal horizons in String Theory
necessitates introduction of additional fields. It has been known for a long time
that charged matter generically destabilizes near-extremal black branes
at a classical level \cite{Gubser:2009qm}. The authors of \cite{Gladden:2024ssb} identified a new classical mechanism
for the instability: the fluctuations of the 'dormant' gauge fields, required by the string theory embedding,
but not activated at the background level.
There is clearly a lot of similarities between the extremal horizons in $\caln=4$ model
and the conifold gauge theory; it would be interesting to understand the genericity
of these similarities.

\section{Technical details}\label{tech}

The relevant effective action is the $\caln=2$ consistent subtruncation of the supersymmetric consistent
truncation of type IIB supergravity on warped deformed conifold with fluxes \cite{Cassani:2010na}:
\begin{equation}
S_{eff}=\frac{1}{2\kappa_5^2} \int_{\calm_5} R \star 1+S_{kin,scal}+S_{kin,vect}+S_{top}+S_{pot}\,, 
\eqlabel{seff}
\end{equation}
with 
\begin{equation}
\begin{split}
&S_{kin,scal}=-\frac{1}{2\kappa_5^2}\int_{\calm_5}\biggl\{
\frac{28}{3} du^2+\frac 43 dv^2+\frac 83 dudv +4dw^2+\frac 12 d\phi^2+\frac 12 e^{2\phi} dC_0^2
\\&+e^{-4u-\phi}|h_1^\Om|^2+e^{-4u+\phi}|g_1^\Om|^2+2 e^{-8 u}f_1^2
\biggr\} \star 1\,,
\end{split}
\eqlabel{sks}
\end{equation}
\begin{equation}
\begin{split}
&S_{kin,vect}=-\frac{1}{2\kappa_5^2}\int_{\calm_5}\biggl\{
\frac 12 e^{\ft 83 u +\ft 83 v}\ (dA)^2+
e^{-\ft 43 u -\ft 43 v }\cosh(2w)\ \biggl[ (da_1^J)^2
+(da_1^\Phi)^2\\&-2\tanh(4w) da_1^J\ \lrcorner\  d a_1^\Phi
\biggr]
\biggr\} \star 1\,,
\end{split}
\eqlabel{svect}
\end{equation}
\begin{equation}
\begin{split}
&S_{top}=\frac{1}{2\kappa_5^2}\int_{\calm_5}\ \left(
A\wedge da_1^J\wedge da_1^J
-A\wedge da_1^\Phi\wedge da_1^\Phi
\right)\,,
\end{split}
\eqlabel{stop}
\end{equation}
\begin{equation}
\begin{split}
&S_{pot}=\frac{1}{2\kappa_5^2}\int_{\calm_5}\biggl\{
24 e^{-\ft {14}{3}u-\ft 23 v}\cosh(2w)-4 e^{-\ft {20}{3}u+\ft 43 v}\cosh(4w)
-2e^{-\ft{32}{3}u-\ft 83 v} f_0^2\\
&-e^{-\ft {20}{3}u-\ft 83 v}\biggl[e^{-\phi}|h_0^\Om|^2+e^\phi |g_0^\Om|^2\biggr]
\biggr\} \star 1\,.
\end{split}
\eqlabel{spot}
\end{equation}
Various fields in \eqref{seff} uplifts to 10d type IIB supergravity as follows:
\begin{itemize}
\item The 10d Einstein frame metric is a direct warped product of metric on $\calm_5$ and
a metric on the deformed coset
\begin{equation}
T^{1,1}\equiv \frac{SU(2)\times SU(2)}{U(1)}\,,
\eqlabel{deft11}
\end{equation}
parameterized by three 0-forms $\{u,v,w\}$ and a single 1-form $A$ on $\calm_5$,
\begin{equation}
\begin{split}
&ds_{10}^2=e^{-\frac 83 u-\frac 23 v}\cdot \underbrace{\quad ds_5^2\quad }_{{\rm metric\ on\ }\ \calm_5}\ +
\underbrace{\sum_{I=1}^5 E^I E^I}_{{\rm metric\ on\ }\ T^{1,1}}\,,\\
&\sum_{I=1}^5 E^I E^I\equiv \frac16 e^{2u+2w}\biggl(e_1^2+e_2^2\biggr)+\frac16 e^{2u-2w}\biggl(e_3^3+e_4^2\biggr)
+\frac19e^{2v}(e_5-3A)^2\,,
\end{split}
\eqlabel{n10d}
\end{equation}
where $e_i$ are the standard coframe 1-forms on $T^{1,1}$ \cite{Minasian:1999tt},
\begin{equation}
\begin{split}
&e_1=-\sin\theta_1\ d(\phi_1)\,,\qquad e_2=d(\theta_1)\,,\\
&e_3=\cos\psi\ \sin\theta_2\ d(\phi_2)-\sin\psi\ d(\theta_2)\,,\\
&e_4=\sin\psi\ \sin\theta_2\ d(\phi_2)+\cos\psi\ d(\theta_2)\,,\\
&e_5=d(\psi)+\cos\theta_1\ d(\phi_1)+\cos\theta_2 d(\phi_2)\,, 
\end{split}
\eqlabel{coframe}
\end{equation} 
for angular coordinates $\{\theta_1,\phi_1,\theta_2,\phi_2,\psi\}$ with ranges
$0\le \theta_{1,2}<\pi$, $0\le \phi_{1,2}<2\pi$, and $0\le \psi <4\pi$.
\item $\phi$ and $C_0$ are type IIB dilaton and axion.
\item To parameterize NSNS and RR 3-form fluxes we introduce left-invariant 1- and 2-forms
on the coset,
\begin{equation}
\begin{split}
&\eta=-\frac 13 e_5\,,\qquad \Omega=\frac 16 (e_1+i e_2)\wedge (e_3-i e_4)\,.\\
\end{split}
\eqlabel{lif}
\end{equation}
\nxt NSNS 2-form potential $B_2$ and a 3-form flux $H_3$ are parameterized by a complex 0-form $b^\Omega$ on $\calm_5$:
\begin{equation}
\begin{split}
H_3=&d(B_2)\,,\qquad B_2=\Re(b^\Om\Om)\,.
\end{split}
\eqlabel{defb2h3}
\end{equation}
The field strength $H_3$ can be decomposed in a basis of left-invariant forms on $T^{1,1}$ \eqref{lif}:
\begin{equation}
\begin{split}
H_3=&\Re\left[h_1^\Om\wedge \Om+h_0^\Om\ \Om\wedge (\eta+A)\right]\,,
\end{split}
\eqlabel{h3decompose}
\end{equation}
where we defined
\begin{equation}
\begin{split}
&h_1^\Om=db^\Om-3i\ A\ b^\Om\equiv Db^\Om\,,\qquad h_0^\Om=3 i\ b^\Om\,.
\end{split}
\eqlabel{shdef}
\end{equation}
\nxt RR 2-form potential $C_2$ and a 3-form flux $F_3$ are parameterized by
a complex 0-form $c^\Omega$ on $\calm_5$:
\begin{equation}
\begin{split}
F_3=d(C_2)-C_0 H_3\,,\qquad C_2=\Re(c^\Om\Om)\,.
\end{split}
\eqlabel{defc2g3}
\end{equation}
The field strength $F_3$ can be decomposed in a basis of left-invariant forms on $T^{1,1}$ \eqref{lif}:
\begin{equation}
\begin{split}
F_3=&\Re\left[g_1^\Om\wedge \Om+g_0^\Om\ \Om\wedge (\eta+A)\right]\,,
\end{split}
\eqlabel{g3decompose}
\end{equation}
where we defined 
\begin{equation}
\begin{split}
&g_1^\Om=dc^\Om-3i\ A\ c^\Om-C_0 Db^\Om\equiv Dc^\Om-C_0 Db^\Om\,,\qquad g_0^\Om=3 i\ (c^\Om-C_0 b^\Om)\,.
\end{split}
\eqlabel{sgdef}
\end{equation}
\item The remaining fields of the effective action \eqref{seff}, \ie the 0-form $f_0$ (in $S_{pot}$,
see \eqref{spot})
\begin{equation}
\begin{split}
f_0=k+3 \Im\left[b^\Om\overline{c^\Om}\right]\,,
\end{split}
\eqlabel{f0def}
\end{equation}
and the 1-form $f_1$ (in $S_{kin,scal}$,
see \eqref{sks})
\begin{equation}
\begin{split}
f_1&=d(a)-2a_1^J-kA+\frac 12 \biggl[
\Re\left[b^\Om\overline{Dc^\Om}\right]-b\leftrightarrow c
\biggr]\,,
\end{split}
\eqlabel{f1def}
\end{equation}
are parameterized by a constant $k$, and additional  0-form $a$ and a 1-form $a_1^J$.
$f_0$, $f_1$ and the remaining 1-form $a_1^\Phi$ describe the self-dual 5-form field strength of type IIB
supergravity\footnote{For detailed discussion see section 2.4 of \cite{Buchel:2014hja}.}.
\item The 5d gravitational coupling $\kappa_5$ is related to the Klebanov-Witten gauge theory central charge $c_{KW}$
as follows:
\begin{equation}
\kappa_5^2=\frac{\pi^2}{c_{KW}}\,,\qquad c_{KW}=\frac{27}{64}N^2\,.
\eqlabel{ccharge}
\end{equation}
\end{itemize}

10d type IIB supergravity is invariant under $SL(2,\reals)$ duality transformation.
This duality is inherited by the consistent truncation \eqref{seff}. As a result, we
can always work in a duality frame with
\begin{equation}
C_0\equiv 0\,,
\eqlabel{c0zero}
\end{equation}
which we will do from now on. The last step is to turn off 3-form fluxes:
\begin{equation}
b^\Om=c^\Om=0\,,
\eqlabel{flux}
\end{equation}
which is a consistent truncation; which in turn implies that the dilaton is constant ---
without loss of generality we set from now on
\begin{equation}
\phi=0\,.
\eqlabel{dil}
\end{equation}
Effective action \eqref{seff} is invariant under the 1-form gauge transformations
(with the 0-form gauge parameters $\alpha\,, \beta$ and $\gamma$):
\begin{equation}
\begin{split}
(I):\ &A\to A+d\alpha\,,\qquad a\to a+k\alpha\,,\\
(II):\ &a_1^J\to a_1^J+d\beta\,,\qquad  a\to a+2\beta\,,\\
(III):\ &a_1^\Phi\to a_1^\Phi+d\gamma\,.
\end{split}
\eqlabel{gt}
\end{equation}
Note that using the bulk gauge transformation we can set $a=0$.
We will keep the scalar $a$ for now, and later fix the gauge transformations
as convenient.

As explained in \cite{Buchel:2024phy}, setting a constant $k=2$ sets the asymptotic
$AdS_5$ radius $L=1$. Furthermore, it is convenient to introduce the linear combinations of the
vector fields as
\begin{equation}
A=\frac 13 \cala+\frac 23 \calv\,,\qquad a_1^J=\frac 13 \calv-\frac 13\cala\,,
\eqlabel{trunc14}
\end{equation}
so that $\cala$ is a graviphoton of  the minimal $\caln=2$ gauged supergravity from the
consistent truncation of type IIB theory
on Sasaki-Einstein manifolds with 5-form flux \cite{Buchel:2006gb}, and $\calv$
is a massive vector dual to a gauge theory operator $\calo_\calv$ of conformal dimension 7.
Finally, the graviphoton is dual to the conserved $R$-symmetry current of the global $U(1)_R$,
and the massless bulk vector $a_1^\Phi$ is dual to the conserved current of the
baryonic $U(1)_B$ symmetry.

To summarize, the effective action $S_{5}$ dual to Klebanov-Witten $\caln=1$ SCFT, which is relevant to study
of its thermal equilibrium states with finite $U(1)_B$ baryonic chemical potential is
a functional of a 5d metric, 3 vector fields, and 4 real scalar
fields\footnote{As it was pointed out in \cite{Cassani:2010na}, this is {\it A More General
Consistent Truncation} of section 7.1 of \cite{Herzog:2009gd}.}:
\begin{equation}
S_{5}\biggl[g_{\mu\nu}\,;\ \underbrace{u,v,w}_{\rm geometry},a\,;\ a_1^\Phi,\cala,\calv\biggr]
=S_{eff}\bigg|_{b^\Om=c^\Om=\phi\equiv 0\,, k=2}.
\eqlabel{skw1}
\end{equation}
The HKPT effective action $S_{HKPT}$, used to construct
baryonic black branes in \cite{Herzog:2009gd}, is a further consistent truncation of \eqref{skw1}:
\begin{equation}
\begin{split}
&S_{HKPT}\biggl[g_{\mu\nu}\,;\ \underbrace{u,v}_{\rm geometry}\,;\ a_1^\Phi\biggr]
=S_{5}\bigg|_{\cala=\calv\equiv 0\,, w=a\equiv 0}\\
&=\frac{1}{2\kappa_5^2}\int_{\calm_5} \biggl\{R-
\frac{28}{3} du^2-\frac 43 dv^2-\frac 83 dudv
-e^{-\ft 43 u -\ft 43 v } (da_1^\Phi)^2-V(u,v)\biggr\}\star 1\,,\\
&V(u,v)=-24 e^{-\ft {14}{3}u-\ft 23 v}+4 e^{-\ft {20}{3}u+\ft 43 v}
+8 e^{-\ft{32}{3}u-\ft 83 v}\,. 
\end{split}
\eqlabel{hkpt}
\end{equation}
Eq.~\eqref{hkpt} reproduces the effective HKPT Lagrangian \cite{Herzog:2009gd},
\begin{equation}
\begin{split}
&\call_{eff}=R-\frac 14 e^{-\ft 43 \chi+2\eta} (F_{\mu\nu}^{HKPT})^2-5(\del_\mu \eta)^2-\frac {10}{3}(\del_\mu\chi)^2
-V^{HKPT}(\eta,\chi)\,,\\
&V^{HKPT}(\eta,\chi)=8 e^{-\ft{20}{3}\chi}+4 e^{-\ft 83\chi}\ \left(e^{-6\eta}-6e^{-\eta}\right)\,,
\end{split}
\eqlabel{leff}
\end{equation}
provided we identify
\begin{equation}
\eta=\frac 25(u-v)\,,\qquad \chi=\frac 25(v+4 u)\,,\qquad \frac{A_\mu^{HKPT}}{\sqrt 2}=a_{1,\mu}^\Phi\,.
\eqlabel{sqrt21}
\end{equation}
The factor of $\sqrt{2}$ difference in the normalization of the HKPT bulk gauge field $F^{HKPT}$ and
$da_1^\Phi$ implies that baryonic chemical potential $\mu^{HKPT}$ used in \cite{Herzog:2009gd}
is related to the corresponding chemical potential $\mu$ used here (see \eqref{uvgen3}) as
\begin{equation}
\mu^{HKPT}=\sqrt 2\ \mu\,.
\eqlabel{sqrt2}
\end{equation}

\subsection{HKPT baryonic black branes}\label{hkpts}

In this section we discuss the thermodynamics of the HKPT baryonic black branes. 
To consider $U(1)_B$-charged black branes in $S_{HKPT}$ \eqref{hkpt}, which realize the gravitational dual
to equilibrium thermal states of the KW gauge theory plasma at finite baryonic chemical potential,
we take the following ansatz:
\begin{equation}
\begin{split}
&ds_5^2=g_{\mu\nu}dx^\mu dx^\nu= -\hat{c}_1^2\ dt^2+\hat{c}_2^2\ d\bm{x}^2+\hat{c}_3^2\ dr^2\,,\qquad \hat{c}_i
=\hat{c}_i(r)\,,\\
&a_1^\Phi=\Phi(r)\ dt+a_{1,r}^\Phi(r)\ dr\,,\qquad \{u,v\}=\{u,v\}(r)\,.
\end{split}
\eqlabel{backw1}
\end{equation}
Using the gauge transformation $(III)$ in \eqref{gt} we can set $a_{1,r}^\Phi\equiv 0$.

It is convenient to introduce
\begin{equation}
\begin{split}
&v\equiv \ln f_1\,,\qquad u\equiv \ln f_2\,, \\
&\hat{c_1}\equiv c_1\ f_1^{1/3}f_2^{4/3}\,,\qquad \hat{c_2}\equiv c_2\ f_1^{1/3}f_2^{4/3}\,,\qquad
\hat{c_3}\equiv c_3\ f_1^{1/3}f_2^{4/3}\,,\\
&c_1=\frac{\sqrt f}{\sqrt r}\,,\qquad c_2=\frac {1}{\sqrt r}\,,\qquad c_3=\frac{s}{2r\sqrt r}\,.
\end{split}
\eqlabel{backw3}
\end{equation}

From \eqref{hkpt} we obtain the following equations of motion:
\begin{equation}
\begin{split}
&0=f_1''-\frac{9r}{16f_2^4 f f_1} (\Phi')^2-\frac{3f_1}{4f_2^2} (f_2')^2
+\frac{7f_1'f_2'}{2f_2}+\left(\frac{f_1}{r f_2}-\frac{f_1 f'}{4f f_2}\right) f_2'
+\biggl(
\frac{15f'}{16f}-\frac{s'}{s}\\&-\frac{3}{4 r}
\biggr) f_1'+\frac{1}{32f r^2 f_2^8 f_1} \biggl(
3 f' f_2^8 f_1^2 r-6 f_2^8 f_1^2 f+12 s^2 f_2^6 f_1^2-34 s^2 f_2^4 f_1^4+28 s^2\biggr)\,,
\end{split}
\eqlabel{bac1}
\end{equation}
\begin{equation}
\begin{split}
&0=f_2''+\frac{9}{4 f_2} (f_2')^2+\left(
\frac{3f'}{4f}-\frac{s'}{s}+\frac{f_1'}{2f_1}\right) f_2'
+\left(\frac{f_2}{4r f_1}-\frac{f_2 f'}{16f f_1}\right) f_1'
-\frac{r}{16f_2^3 f f_1^2} (\Phi')^2\\
&+\frac{1}{32f r^2 f_2^7 f_1^2}
\biggl(3 f' f_2^8 f_1^2 r-6 f_2^8 f_1^2 f-36 s^2 f_2^6 f_1^2+14 s^2 f_2^4 f_1^4+28 s^2\biggr)\,,
\end{split}
\eqlabel{bac2}
\end{equation}
\begin{equation}
\begin{split}
&0=\Phi''-\left(\frac{s'}{s}+\frac{f_1'}{f_1}\right)\Phi'\,,
\end{split}
\eqlabel{bac3}
\end{equation}
\begin{equation}
\begin{split}
&0=f'+\frac{2}{r f_1 f_2^7 (-2 f_2 f_1' r-8 f_1 f_2' r+3 f_2 f_1)}
\biggl(-8 f_2^7 f_1 f_2' f f_1' r^2-12 f_2^6 f_1^2 (f_2')^2 f r^2
\\&+4 f_2^8 f_1 f f_1' r+16 f_2^7 f_1^2 f_2' f r-3 f_2^8 f_1^2 f+6 s^2 f_2^6 f_1^2
-s^2 f_2^4 f_1^4-(\Phi')^2 f_2^4 r^3-2 s^2\biggr)\,,
\end{split}
\eqlabel{bac4}
\end{equation}
\begin{equation}
\begin{split}
&0=s'+\frac{s}{f_2^8 f_1^2 f r (-2 f_2 f_1' r-8 f_1 f_2' r+3 f_2 f_1)}
\biggl(
2 f_2^9 f_1 f (f_1')^2 r^2+8 f_2^7 f_1^3 (f_2')^2 f r^2\\&+f_2^9 f_1^2 f f_1' r
+4 f_2^8 f_1^3 f_2' f r+12 s^2 f_2^7 f_1^3-2 s^2 f_2^5 f_1^5-2 (\Phi')^2 f_2^5 f_1' r^4
-8 (\Phi')^2 f_2^4 f_1 f_2' r^4\\&+(\Phi')^2 f_2^5 f_1 r^3
+4 s^2 f_2 f_1' r+16 s^2 f_1 f_2' r-10 s^2 f_2 f_1
\biggr)\,.
\end{split}
\eqlabel{bac5}
\end{equation}
The equations of motion \eqref{bac1}-\eqref{bac5}
describing baryonic black branes are solved subject to the following
asymptotics:
\nxt in the UV, \ie as $r\to 0_+$,
\begin{equation}
\begin{split}
&f=1+ f_4\ r^2+\frac23 a_2^2\ r^3 +\calo(r^5)\,,
\end{split}
\eqlabel{uvgen1}
\end{equation}
\begin{equation}
\begin{split}
&s=1+\frac{7}{30} a_2^2\ r^3-15 f_{1,8}\ r^4 +\calo(r^5)\,,
\end{split}
\eqlabel{uvgen2}
\end{equation}
\begin{equation}
\begin{split}
&\Phi=\mu+a_2\ r+\left(
\frac{5}{96} a_2^3+\frac14 a_2 f_{1,6}+\frac{1}{40} a_2^3\ \ln r
\right)\ r^4
+\calo(r^5)\,,
\end{split}
\eqlabel{uvgen3}
\end{equation}
\begin{equation}
\begin{split}
&f_1=1+\left(f_{1,6}+\frac{1}{10} a_2^2\ \ln r\right)\ r^3+f_{1,8}\ r^4+\calo(r^5\ln r)\,,
\end{split}
\eqlabel{uvgen5}
\end{equation}
\begin{equation}
\begin{split}
&f_2=1+\left(
-\frac14 f_{1,6}-\frac{1}{40} a_2^2-\frac{1}{40} a_2^2\ \ln r\right)\ r^3+ f_{1,8}\ r^4+\calo(r^5\ln r)\,,
\end{split}
\eqlabel{uvgen6}
\end{equation}
specified by 
\begin{equation}
\biggl\{\
a_2\,,\, f_4\,,\, f_{1,6}\,,\,  f_{1,8}
\
\biggr\}\,,
\eqlabel{uvpar}
\end{equation}
as functions of a $U(1)_B$ chemical potential $\mu$; 
\nxt in the IR, \ie as $y\equiv 1-r\to 0_+$,
\begin{equation}
\begin{split}
&f_1=f_{1,0}^h+\calo(y)\,,\qquad f_2=f_{2,0}^h+\calo(y)\,,\qquad  s=s^h_0+\calo(y)\,,\\
&
\Phi=a_1^h\ y+\calo(y^2)\,,\qquad f=\frac{2(s_0^h)^2-(a^h_1)^2(f_{2,0}^h)^4}{(f_{2,0}^h)^8 (f_{1,0}^h)^2}\ y+\calo(y^2)\,,
\end{split}
\eqlabel{irass}
\end{equation}
specified by 
\begin{equation}
\biggl\{\
s^h_0\,,\, f_{1,0}^h\,,\, f^h_{2,0}\,,\, a_1^h
\
\biggr\}\,,
\eqlabel{irpar}
\end{equation}
again, as functions of a $U(1)_B$ chemical potential $\mu$. 
Note that in total, we have 4 parameters in the UV \eqref{uvpar}
and 4 parameters in the IR \eqref{irpar}, precisely as needed to
specify a solution of a coupled system of 3 second order ODEs \eqref{bac1}-\eqref{bac3},
and a pair of first order ODEs \eqref{bac4} and \eqref{bac5}, given a value of the
baryonic chemical potential $\mu$.

In practice, we solve equations of motion for $f,s,\Phi,f_1,f_2$ 
using the shooting method codes adopted from \cite{Aharony:2007vg}.
Once baryonic black brane solutions are constructed, we can use the holographic
renormalization\footnote{In this model a simple holographic renormalization of
\cite{Balasubramanian:1999re}
is enough.}
to extract their thermodynamic properties:
\begin{equation}
\begin{split}
&2\pi T=\frac{2(s_0^h)^2-(a^h_1)^2(f_{2,0}^h)^4}{(f_{2,0}^h)^8 (f_{1,0}^h)^2s_0^h}\,,\qquad
\hat\cale\equiv 2\kappa_5^2 \cale =-3 f_4\,,\qquad \hat\rho_B\equiv  2\kappa_5^2 \rho_B =-4 a_2\,,
\\
&\hat{\cals}\equiv 2\kappa_5^2 \cals =4\pi f^h_{1,0} (f^h_{2,0})^4\,,\qquad
\hat\Omega\equiv 2\kappa_5^2 \Omega =4 \mu a_2 -3 f_4 -T\hat\cals\,,
\end{split}
\eqlabel{thermo}
\end{equation}
where $T$ is the temperature,  $\cals$ is the entropy density, $\Omega$ is the Gibbs free energy density,
$\cale$ is the energy density and $\rho_B$ is the $U(1)_B$ symmetry charge density.
The thermodynamics of  baryonic black branes is reviewed in section \ref{phase}.

An important check on numerics is provided by enforcement of
the first law of thermodynamics (A), and the conformality of the
black brane phase (B):
\begin{equation}
\begin{split}
&(A):\qquad d\hat\Omega=-\hat\cals dT-\hat\rho_B d\mu\qquad \Longrightarrow\qquad 0=\frac{\mu^3}{\hat\cals}\cdot \left(\frac{\del \frac{\hat\Omega}{\mu^4}}{\del \frac T\mu}\right)\bigg|_{\mu={\rm const}} +1\,,\\
&(B):\qquad \hat\cale=3 P=-3 \hat\Omega\qquad \Longrightarrow\qquad
0=\frac{\hat{\cale}}{3\hat{\Omega}}+1\,.
\end{split}
\eqlabel{thermconst}
\end{equation}
We find that the constraint (A) is satisfied at the level $\propto 10^{-7}$, and the constraint
$(B)$ is satisfied at the level $\propto 10^{-11}$.

\subsection{Hydrodynamics of baryonic black branes}\label{hydro}

In this section we demonstrate that the HKPT baryonic black branes \cite{Herzog:2009gd}
(also reviewed in section \ref{hkpts} ) become dynamically unstable to spatial
'clumping' of the $U(1)_R$ symmetry charge below the critical temperature \eqref{tmuc}.

Within the effective action \eqref{skw1} we consider linearized fluctuations 
\begin{equation}
\begin{split}
&\cala=\delta\cala_t\ dt+\delta\cala_z dz+\delta\cala_r\ dr\,,  \calv
=\delta\calv_t\ dt+\delta\calv_z dz+\delta\calv_r\ dr\,, w=\delta w\,, a=\delta a\,,\\
&\delta\cala_{t,z,r}=e^{-i \omega t +i q z}\cdot \delta\cala_{t,z,r}(r)\,,\qquad
\delta w=e^{-i \omega t +i q z}\cdot \delta W(r)\,,\qquad \delta a=e^{-i \omega t +i q z}\cdot \delta a(r)\,,
\end{split}
\eqlabel{fluc}
\end{equation}
about the baryonic black brane background \eqref{backw1}.
It is straightforward to verify that the set \eqref{fluc} will decouple from the
remaining fluctuations in the helicity-0 (the sound channel) sector. We use the bulk gauge
transformations \eqref{gt} to set
\begin{equation}
\delta \cala_r=\delta\calv_r \equiv 0\,.
\eqlabel{gfixdiff}
\end{equation}
The equations of motion for the remaining fluctuations take the form\footnote{The background warp factors $c_i$ are parameterized as in \eqref{backw3}.}:
\begin{equation}
\begin{split}
&0={\delta\cala_t}''+\biggl(
\frac{f_1'}{3f_1}+\frac{3c_2'}{c_2}+\frac{4f_2'}{3f_2}-\frac{c_3'}{c_3}-\frac{c_1'}{c_1}\biggr)\
{\delta\cala_t}'
+\frac 83\biggl(
\frac{f_1'}{f_1}+\frac{f_2'}{f_2}\biggr)\ {\delta\calv_t}'
+8 \Phi'\ {\delta W}'\\
&-\frac{c_3^2 q }{c_2^2}\ (q\ {\delta\cala_t} +\omega\ {\delta\cala_z} )+ \frac{8(f_1^4 f_2^4-1) c_3^2}{f_1^2 f_2^8}\  (i {\delta a}\ \omega+2\ {\delta\calv_t})\,,
\end{split}
\eqlabel{fl1}
\end{equation}
\begin{equation}
\begin{split}
&0={\delta\cala_z}''+\biggl(
-\frac{c_3'}{c_3}+\frac{c_2'}{c_2}+\frac{c_1'}{c_1}+\frac{f_1'}{3f_1}+\frac{4f_2'}{3f_2}
\biggr)\ {\delta\cala_z}'
+\frac 83 \biggl(\frac{f_1'}{f_1}+\frac{f_2'}{f_2}\biggr)\ {\delta\calv_z}'\\
&-\frac{8(f_1^4 f_2^4-1) c_3^2}{f_1^2 f_2^8}
\ (i{\delta a}\ q-2\ {\delta\calv_z})+\frac{c_3^2 \omega}{c_1^2}\ ({\delta\cala_t}\ q+{\delta\cala_z}\ \omega)\,,
\end{split}
\eqlabel{fl2}
\end{equation}
\begin{equation}
\begin{split}
&0={\delta\calv_t}''+\biggl(
\frac{5f_1'}{3f_1}+\frac{8f_2'}{3f_2}-\frac{c_3'}{c_3}-\frac{c_1'}{c_1}+\frac{3 c_2'}{c_2}
\biggr)\ {\delta\calv_t}'
+\frac 43\biggl (\frac{f_1'}{f_1}+\frac{f_2'}{f_2}\biggr)\ {\delta\cala_t}'-4 \Phi'\ {\delta W}'\\
&-\frac{4  (f_1^4 f_2^4+2) c_3^2}{f_1^2 f_2^8}\ (i {\delta a}\ \omega+2\ {\delta\calv_t})
-\frac{c_3^2 q} {c_2^2}\ (q\ {\delta\calv_t}+\omega\ {\delta\calv_z})\,,
\end{split}
\eqlabel{fl3}
\end{equation}
\begin{equation}
\begin{split}
&0={\delta\calv_z}''+\biggl(
\frac{c_1'}{c_1}+\frac{5f_1'}{3f_1}+\frac{8f_2'}{3f_2}-\frac{c_3'}{c_3}+\frac{c_2'}{c_2}
\biggr)\  {\delta\calv_z}'
+\frac 43 \biggl( \frac{f_1'}{f_1}+\frac{f_2'}{f_2}\biggr)\ {\delta\cala_z}'
\\&+\frac{4 (f_1^4 f_2^4+2) c_3^2}{f_1^2 f_2^8}\  (i{\delta a}\ q-2\ {\delta\calv_z})
+\frac{c_3^2 \omega} {c_1^2}\ (q\ {\delta\calv_t}+\omega\ {\delta\calv_z})\,,
\end{split}
\eqlabel{fl4}
\end{equation}
\begin{equation}
\begin{split}
&0={\delta W}''+\biggl(
\frac{4 f_2'}{f_2}-\frac{c_3'}{c_3}+\frac{c_1'}{c_1}+\frac{3c_2'}{c_2}+\frac{f_1'}{f_1}
\biggr)\ {\delta W}' +\frac{\Phi'}{3f_1^2 c_1^2 f_2^4} ({\delta\cala_t}'-{\delta\calv_t}')
\\&+ \biggl(
\frac{2(\Phi')^2}{f_1^2 c_1^2 f_2^4}
+\frac{c_3^2 (f_2^4 \omega^2-4 c_1^2 (2 f_1^2-3 f_2^2))}{c_1^2 f_2^4}-\frac{c_3^2 q^2}{c_2^2}\biggr)
\ {\delta W}\,,
\end{split}
\eqlabel{fl5}
\end{equation}
\begin{equation}
\begin{split}
&0={\delta a}''+\biggl(
\frac{3c_2'}{c_2}-\frac{c_3'}{c_3}-\frac{4 f_2'}{f_2}+\frac{c_1'}{c_1}+\frac{f_1'}{f_1}
\biggr)\  {\delta a}'-\frac{\omega c_3^2} {c_1^2}\ (2 i\ {\delta\calv_t}-{\delta a}\ \omega)
\\&-\frac{c_3^2 q}{c_2^2}\  ({\delta a}\ q+2 i\ {\delta\calv_z})\,,
\end{split}
\eqlabel{fl6}
\end{equation}
along with the gauge fixing constraints from \eqref{gfixdiff}:
\begin{equation}
\begin{split}
&0=\frac \omega3 c_2^2 f_2^4\ \biggl(
12\Phi'\  {\delta W}+{\delta\cala_t}'
-{\delta\calv_t}'\biggr)
+\frac q3 c_1^2 f_2^4\ \biggl(
{\delta\cala_z}'-{\delta\calv_z}'\biggr)
-4 i c_2^2  f_1^2 c_1^2\ {\delta a}'\,,
\end{split}
\eqlabel{flc1}
\end{equation}
\begin{equation}
\begin{split}
&0=\frac \omega3 c_2^2 f_1^2 f_2^8\ \biggl(
{\delta\cala_t}'+2\ {\delta\calv_t}'
\biggr)+\frac q3 c_1^2 f_1^2 f_2^8\ \biggl(
{\delta\cala_z}'+2\ {\delta\calv_z}'
\biggr)+8 i c_2^2 c_1^2\ \delta a'\,.
\end{split}
\eqlabel{flc2}
\end{equation}
We explicitly verified that \eqref{flc1} and \eqref{flc2} are consistent with \eqref{fl1}-\eqref{fl6}.

Further introducing
\begin{equation}
Z_0\equiv q\ \delta\cala_t+\omega\ \delta\cala_z\,,\qquad
Z_1\equiv q\ \delta\calv_t+\omega\ \delta\calv_z\,,\qquad Z_a\equiv \delta a'\,,\qquad Z_w\equiv \delta W\,,
\eqlabel{flvar}
\end{equation}
we find from \eqref{fl1}-\eqref{flc2}:
\begin{equation}
\begin{split}
&0=Z_0''+\biggl(
\frac{(c_2^2 \omega^2+c_1^2 q^2) c_1'}{c_1 (c_2^2 \omega^2-c_1^2 q^2)}
+\frac{(c_2^2 \omega^2-3 c_1^2 q^2) c_2'}{c_2 (c_2^2 \omega^2-c_1^2 q^2)}
-\frac{c_3'}{c_3}+\frac{f_1'}{3f_1}
+\frac{4f_2'}{3f_2}\biggr)\ Z_0'
\\&+\frac 83 \biggl(
\frac{f_1'}{f_1}+\frac{f_2'}{f_2}\biggr)\  Z_1'+8 q \Phi'\ Z_w'
+\frac{c_3^2 (c_2^2 \omega^2-c_1^2 q^2)}{c_2^2 c_1^2}\ Z_0+\frac{16 c_3^2 (f_1^4 f_2^4-1)}{f_2^8 f_1^2}\ Z_1
\\&+\frac{16 (c_2 c_1'-c_1 c_2') c_2 \omega q}{c_1 f_1^2 f_2^8 (c_2^2 \omega^2-c_1^2 q^2)}\
\biggl(-i c_1^2 (f_1^4 f_2^4-1)\ Z_a+f_1^2 \Phi' f_2^8 \omega\ Z_w\biggr)\,,
\end{split}
\eqlabel{ffl1}
\end{equation}
\begin{equation}
\begin{split}
&0=Z_1''+\biggl(
\frac{(c_2^2 \omega^2+c_1^2 q^2)c_1'}{c_1 (c_2^2 \omega^2-c_1^2 q^2)}
+\frac{(c_2^2 \omega^2-3 c_1^2 q^2) c_2'}{c_2 (c_2^2 \omega^2-c_1^2 q^2)}
-\frac{c_3'}{c_3}+\frac{5f_1'}{3f_1}+\frac{8f_2'}{3f_2}\biggr)\ Z_1'
\\&+\frac 43 \biggl( \frac{f_1'}{f_1}+\frac{f_2'}{f_2}\biggr)\ Z_0'
-4 q \Phi'\ Z_w'+c_3^2 \biggr(
\frac{c_2^2 \omega^2-c_1^2 q^2}{c_2^2 c_1^2}-\frac{8 (f_1^4 f_2^4+2)}{f_1^2 f_2^8}\biggr)\ Z_1
\\&-\frac{8 (c_2 c_1'-c_1 c_2') c_2 \omega q}{c_1 f_1^2 f_2^8 (c_2^2 \omega^2-c_1^2 q^2)}
\biggl( -i  c_1^2 (f_1^4 f_2^4+2)\ Z_a+f_1^2 \Phi' f_2^8 \omega\ Z_w\biggr)\,,
\end{split}
\eqlabel{ffl2}
\end{equation}
\begin{equation}
\begin{split}
&0=Z_w''+\biggl(
\frac{4 f_2'}{f_2}-\frac{c_3'}{c_3}+\frac{c_1'}{c_1}+\frac{3c_2'}{c_2}+\frac{f_1'}{f_1}
\biggr)\  Z_w'
+\biggl(
\frac{2(\Phi')^2}{f_1^2 c_1^2 f_2^4}
+\frac{c_3^2 (f_2^4 \omega^2-4 c_1^2 (2 f_1^2-3 f_2^2))}{c_1^2 f_2^4}\\&-\frac{c_3^2 q^2}{c_2^2}\biggr)\ Z_w
-\frac{\Phi'}{3f_1^2 f_2^8 c_1^2 (c_2^2 \omega^2-c_1^2 q^2)}
\biggl(12 c_2^2 \Phi' f_2^4 \omega^2\ Z_w-12 i c_2^2  f_1^2 c_1^2 \omega\ Z_a\\&+c_1^2 f_2^4 q\
(Z_0'-Z_1')\biggr)\,,
\end{split}
\eqlabel{ffl3}
\end{equation}
\begin{equation}
\begin{split}
&0=Z_a''+\biggl(
\frac{(3 c_2^2 \omega^2-c_1^2 q^2) c_1'}{c_1 (c_2^2 \omega^2-c_1^2 q^2)}
+\frac{(3 c_2^2 \omega^2-5 c_1^2 q^2) c_2'}{c_2 (c_2^2 \omega^2-c_1^2 q^2)}
-\frac{3 c_3'}{c_3}+\frac{f_1'}{f_1}-\frac{4 f_2'}{f_2}\biggr)\ Z_a'
+\biggl(
\frac{3 (c_3')^2}{c_3^2}\\
&-\frac{c_3''}{c_3}-\frac{7(\Phi')^2}{16 f_1^2 c_1^2 f_2^4} 
+\frac{c_3^2(c_2^2 \omega^2-c_1^2 q^2)}{c_2^2 c_1^2} +\frac{2 q^2 c_1 (c_2 c_1'
-c_1 c_2')}{c_2 (c_2^2 \omega^2-c_1^2 q^2)} \biggl(
\frac{3 c_2'}{c_2}-\frac{c_3'}{c_3}-\frac{4f_2'}{f_2}
+\frac{c_1'}{c_1}+\frac{f_1'}{f_1}\biggr)
\\&-\frac{3c_1'c_3'}{c_1 c_3}
-\frac{f_1'c_3'}{c_3 f_1}-\frac{3 c_2'c_3'}{c_3 c_2}
+\frac{4 f_2'c_3'}{c_3 f_2}+\frac{(c_1')^2}{c_1^2}
+\frac{c_1' f_1'}{8c_1 f_1}
+\frac{3c_1'c_2'}{8c_1 c_2}
-\frac{15c_1'f_2'}{2c_1 f_2}
-\frac{(f_1')^2}{f_1^2}
-\frac{45c_2'f_1'}{8f_1 c_2}
\\&+\frac{f_2'f_1'}{2f_2 f_1}
-\frac{69(c_2')^2}{8c_2^2}+\frac{3c_2'f_2'}{2f_2 c_2}
+\frac{67(f_2')^2}{4f_2^2}+\frac{17c_3^2 f_1^2}{4f_2^4}-\frac{51c_3^2}{2f_2^2}
+\frac{25c_3^2}{2f_2^8 f_1^2}
\biggr)\ Z_a
-\frac{8 i \Phi' \omega c_3^2}{c_1^2}\ Z_w\\
&-\frac{4 i  c_3^2 \omega q
(c_2 c_1'-c_1 c_2')}{c_2 c_1 (c_2^2 \omega^2-c_1^2 q^2)}\ Z_1\,.
\end{split}
\eqlabel{ffl4}
\end{equation}
Solutions of \eqref{ffl1}-\eqref{ffl4} with appropriate boundary conditions determine
the spectrum of baryonic black branes quasinormal modes --- equivalently
the physical spectrum of linearized fluctuations in conifold gauge theory plasma with
a baryonic chemical potential \eqref{diff}. Following \cite{Kovtun:2005ev,Son:2002sd}
we impose the incoming-wave boundary conditions at the black brane horizon, and 'normalizability'
at asymptotic $AdS_5$ boundary.
Focusing on the $\Re[\ww]=0$  diffusive branch, and
introducing\footnote{The shift of the IR index for $Z_a$ is due to the fact that
$Z_a=\delta a'$, see \eqref{flvar}.}
\begin{equation}
Z_{0,1,w}=\left(\frac{c_1}{c_2}\right)^{-i\ww} z_{0,1,w}\,,\qquad Z_a= \left(\frac{c_1}{c_2}\right)^{-i\ww-2} z_a\,,
\qquad \ww=-i v\ \kk\,,
\eqlabel{income}
\end{equation}
we solve \eqref{ffl1}-\eqref{ffl4} subject to the asymptotics: 
\nxt in the UV, \ie as $r\to 0_+$,
\begin{equation}
\begin{split}
&z_0=\kk\ r+\left(\frac12 \pi^2 T^2 \kk^3 (v^2+1)-4 a_2 z_{w;2}\right)\ r^2+\calo(r^3)\,,\\
&z_1=-a_2 z_{w;2}\ r^2+\left(z_{1;6}+\frac75 a_2 \kk^2 \pi^2 T^2 z_{w;2} (v^2+1) \ln r\right)\ r^3+\calo(r^4\ln r)\,,\\
&z_w=z_{w;2}\ r+\kk^2 \pi^2 T^2 z_{w;2} (v^2+1)\ r^2+\calo(r^3)\,,\\
&z_a=-a_2 v T \pi z_{w;2}\ r^2+\left(z_{a;6}+\frac35 a_2 v T^3 \pi^3 \kk^2 z_{w;2} (v^2+1)\ \ln r\right)\ r^3
+\calo(r^4\ln r)\,,
\end{split}
\eqlabel{uv1}
\end{equation}
specified, for a fixed background and a momentum $\kk$,  by
\begin{equation}
\biggl\{\
v\,,\ z_{w;2}\,,\ z_{1;6}\,,\ z_{a;6}
\
\biggr\}\,;
\eqlabel{fluv}
\end{equation}
\nxt in the IR, \ie as $y\equiv 1-r\to 0_+$,
\begin{equation}
\begin{split}
&z_{0,1,w,a}=z_{0,1,w,a;0}^h\ +\calo(y)\,,
\end{split}
\eqlabel{ir1}
\end{equation}
specified  by
\begin{equation}
\biggl\{\
z_{0;0}^h\,,\ z_{1;0}^h\,,\ z_{w;0}^h\,,\ z_{a;0}^h
\
\biggr\}\,.
\eqlabel{flir}
\end{equation}
Note that in total we have $4+4=8$ parameters, see \eqref{fluv} and \eqref{flir},
which is precisely what is necessary to identify a solution of a coupled system of
4 second-order ODEs \eqref{ffl1}-\eqref{ffl4}. Furthermore, without
the loss of generality we normalized the solutions 
so that
\begin{equation}
\lim_{r\to 0} \frac{dz_0}{dr}=\kk\,.
\eqlabel{normaliz}
\end{equation}

\begin{figure}[ht]
\begin{center}
\psfrag{m}[tt][][1.0][0]{{$\mu$}}
\psfrag{d}[tt][][1.0][0]{{$D$}}
\psfrag{w}[tt][][1.0][0]{{$\Im[\ww]$}}
\psfrag{k}[tt][][1.0][0]{{$\kk$}}
\includegraphics[width=3in]{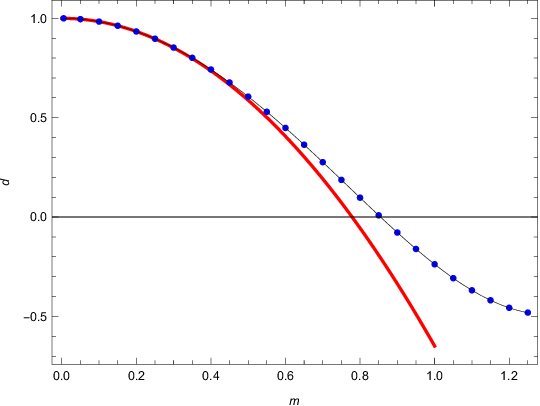}
\includegraphics[width=3in]{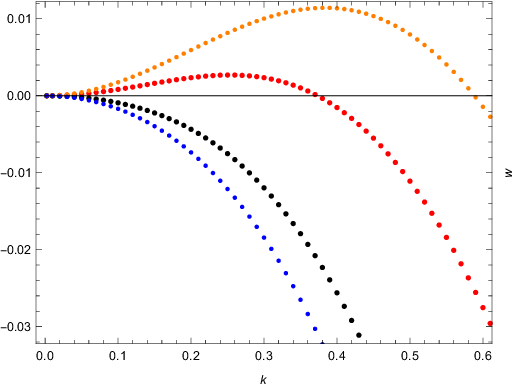}
\end{center}
  \caption{Left panel: the diffusive coefficient $D$ (blue dots, see \eqref{resfl})
  as a function of the baryonic chemical potential $\mu_B$ of HKPT black branes. Its negative values indicate
 R-charge clumping instability of translationary invariant horizons of baryonic black branes.
 Right panel: the imaginary part of the diffusive QNM frequency $\ww=\frac{\omega}{2\pi T}$ as a function of
 a spatial momenta $\kk=\frac{q}{2\pi T}$ for select values of $\frac{T}{\mu_B}$ \eqref{dispsets}.
 For small values of $\kk$, the modes in plasma with $T<T_{crit}$ \eqref{tmuc} (the red and orange sets
 of dots) are unstable. 
} \label{figure2}
\end{figure}

In practice, given a fixed baryonic black brane geometry, numerically
obtained solving \eqref{bac1}-\eqref{bac5} with \eqref{uvgen1}-\eqref{uvgen6} and \eqref{irass},
we solve \eqref{ffl1}-\eqref{ffl4}, subject to the asymptotics \eqref{uv1} and \eqref{ir1}
for a range of momenta $\kk\in [-0.1,0.1]$. Obtained values of $v\equiv v(\kk)$ allows
to compute the dispersion relation for the quasinormal modes, and the diffusion coefficient in \eqref{diff}:
\begin{equation}
\Im[\ww]=-v(\kk)\cdot \kk\,,\qquad D=\frac{dv(\kk)}{d\kk}\bigg|_{\kk=0}\,,
\eqlabel{resfl}
\end{equation}
as  functions of a baryonic chemical potential $\mu$. 
The results are these analyses are shown in fig.\ref{figure2}.
The left panel shows the diffusive coefficient $D$ \eqref{resfl} as a function
of $\mu$ (the blue dots) --- it is positive for small values of $\mu$ (the large values of $\frac {T}{\mu_B}$),
and negative for sufficiently large values of the chemical potential. $D$ vanishes at
\begin{equation}
\mu=\mu_*=0.85501(6)\qquad \Longrightarrow\qquad \frac{T}{\mu_*}=\frac{T}{\mu_B}\bigg|_{crit}=0.2770(5)\,,
\eqlabel{mus}
\end{equation}
corresponding to a critical temperature \eqref{tmuc}. The solid red line represents
the fit to the diffusion coefficient at small $\mu$:
\begin{equation}
D\bigg|_{red}=1-e^{1/2} \mu^2\,.
\eqlabel{dred}
\end{equation}
In the right panel we provide a sample of the dispersion relations for the modes
above the critical temperature (the sets of blue and black dots), and below the critical
temperature (the sets of orange and red dots):
\begin{equation}
\frac{T}{\mu}=\biggl\{\ \mathcolor{orange}{0.8}\,,\ \mathcolor{red}{0.9}\,,\ 
\mathcolor{black}{1.1}\,,\ \mathcolor{blue}{1.2}
\
\biggr\}\cdot \frac{T}{\mu_B}\bigg|_{crit}\,.
\eqlabel{dispsets}
\end{equation}
Analogously to the $\caln=4$ SYM plasma discussed in \cite{Gladden:2024ssb},
we find that the extremal baryonic black brane limit
is never realized --- there is a classical instability (at sufficiently low temperatures)
in the conifold  gauge theory plasma,
associated with the spatial clumping of the $U(1)_R$ symmetry charge, even though the
corresponding equilibrium thermal state has only a baryonic chemical potential/charge.

\subsection{Onset of spatially homogeneous instability of baryonic black branes}\label{onset}

In section \ref{hydro} we established that translationary invariant horizons of
baryonic black branes become unstable below $T_{crit}$ \eqref{tmuc}.
Identical physical phenomenon was demonstrated in $\caln=4$ SYM plasma
in \cite{Gladden:2024ssb}. In the latter case, precisely at $T_{crit}$,
there is also an onset of a new homogeneous phase of SYM plasma\footnote{See also
\cite{Henriksson:2019zph,Anabalon:2024lgp} for earlier discussions.}
\cite{Buchel:2025cve,Buchel:2025tjq}. In this section we establish the existence of a similar
new homogeneous phase of HKPT black branes.    

Since we are interested in potential instabilities leading to a homogeneous phase,
we set $q=0$ in the equations of motion for the linearized fluctuations about the
baryonic black brane background \eqref{fl1}-\eqref{flc1}. This allows for a truncation
with
\begin{equation}
\delta \cala_z=\delta \calv_z\equiv 0\,.
\eqlabel{truncz}
\end{equation}
To proceed, we need to be careful with $\omega\to 0$ limit.
\begin{itemize}
\item {\bf{(O-I)}}: We can directly set $\omega=0$, which requires vanishing of $\delta a$,
in \eqref{fl1}-\eqref{flc1} and search for the normalizable
solutions of the resulting equations (as was done in \cite{Buchel:2025cve} for a SYM),
\begin{equation}
\begin{split}
&0={\delta\cala_t}''+\biggl(
\frac{f_1'}{3f_1}+\frac{3c_2'}{c_2}+\frac{4f_2'}{3f_2}-\frac{c_3'}{c_3}-\frac{c_1'}{c_1}\biggr)\
{\delta\cala_t}'
+\frac 83\biggl(
\frac{f_1'}{f_1}+\frac{f_2'}{f_2}\biggr)\ {\delta\calv_t}'
+8 \Phi'\ {\delta W}'\\
&+\frac{16(f_1^4 f_2^4-1) c_3^2}{f_1^2 f_2^8}\  {\delta\calv_t}\,,
\end{split}
\eqlabel{on1}
\end{equation}
\begin{equation}
\begin{split}
&0={\delta\calv_t}''+\biggl(
\frac{5f_1'}{3f_1}+\frac{8f_2'}{3f_2}-\frac{c_3'}{c_3}-\frac{c_1'}{c_1}+\frac{3 c_2'}{c_2}
\biggr)\ {\delta\calv_t}'
+\frac 43\biggl (\frac{f_1'}{f_1}+\frac{f_2'}{f_2}\biggr)\ {\delta\cala_t}'-4 \Phi'\ {\delta W}'\\
&-\frac{8  (f_1^4 f_2^4+2) c_3^2}{f_1^2 f_2^8}\  {\delta\calv_t}\,,
\end{split}
\eqlabel{on2}
\end{equation}
\begin{equation}
\begin{split}
&0={\delta W}''+\biggl(
\frac{4 f_2'}{f_2}-\frac{c_3'}{c_3}+\frac{c_1'}{c_1}+\frac{3c_2'}{c_2}+\frac{f_1'}{f_1}
\biggr)\ {\delta W}' +\frac{\Phi'}{3f_1^2 c_1^2 f_2^4} ({\delta\cala_t}'-{\delta\calv_t}')
\\&+ \biggl(
\frac{2(\Phi')^2}{f_1^2 c_1^2 f_2^4}
-\frac{4c_3^2  (2 f_1^2-3 f_2^2)}{c_1^2 f_2^4}\biggr)
\ {\delta W}\,.
\end{split}
\eqlabel{on3}
\end{equation}
\item {\bf{(O-II):}} In the limit $\omega\to 0$, but without directly setting $\omega=0$, we can eliminate
$\delta a'$ from \eqref{flc1} and substitute the result into \eqref{flc2}. We find in this case an additional
first-order constraint
on $\{\delta \cala_t, \delta\calv_t, \delta W\}$ fluctuations, namely,
\begin{equation}
0=\omega\cdot \biggl\{\ (f_1^4 f_2^4+2)\ \delta\cala_t'+2 (f_1^4 f_2^4-1)\ \delta\calv_t'+24\Phi'\ \delta W\ \biggr\}\,.
\eqlabel{wconst}
\end{equation}
We kept an overall factor $\omega$ to emphasize that \eqref{wconst} is irrelevant in the
strict limit $\omega=0$. It is straightforward to verify that \eqref{wconst} is consistent
with the second-order equations \eqref{on1}-\eqref{on2}, provided we use the background equation of motion
\eqref{bac3}. 
\end{itemize}

We show now that \eqref{on1}-\eqref{on2} allows for a normalizable solution,
and thus for an onset of a homogeneous phase,
but that this solution is inconsistent with the constraint \eqref{wconst}. 
Physically, this suggests that the homogeneous phase predicted here (it will
be further analyzed in section \ref{normal}) can not arise from
the hydrodynamic instability\footnote{We believe the same is
true for the ordered phase of the SYM plasma \cite{Buchel:2025cve,Buchel:2025tjq}.}. 

The easiest way to identify the onset of an instability for a gravitational mode is to turn on
its source term; the instability is then signalled by the divergence of its normalizable
component\footnote{See for example section 5 of \cite{Buchel:2019pjb}.}.  
Thus, we seek solutions of \eqref{on1}-\eqref{on3} with the following asymptotics:
\nxt In the UV, \ie as $r\to 0_+$,
\begin{equation}
\begin{split}
&\delta W=(\colorbox{yellow}{$z_{w;2}$}+\colorbox{red}{1} \ln r)\ r+\calo(r^3\ln r)\,,\\
&\delta \calv_t=\left(-a_2 z_{w;2}-\frac74 a_2-a_2 \ln r\right)\ r^2+z_{1;6}\ r^3+\calo(r^4\ln r)\,,\\
&\delta\cala_t=z_{0;2}\ r+(-4 a_2 z_{w;2}+2 a_2-4 a_2 \ln r)\ r^2+\calo(r^4\ln r)\,,
\end{split}
\end{equation}
specified by
\begin{equation}
\biggl\{\
z_{w;2}\,,\ z_{0;2}\,,\ z_{1;6}
\
\biggr\}\,.
\eqlabel{susnc}
\end{equation}
We highlighted the source
term $\colorbox{red}{1}$ (which can be set to 1 as the equations are linear), and the
corresponding  normalizable coefficient $\colorbox{yellow}{$z_{w;2}$}$.
\nxt In the IR, \ie as $y\equiv 1-r\to 0_+$,
\begin{equation}
\delta W=z_{w;0}^h +\calo(y)\,,\qquad \delta\cala_t=z_{0;1}^h\ y +\calo(y^2)\,,\qquad
\delta\calv_t=z_{1;1}^h\ y+\calo(y^2)\,,
\eqlabel{susir}
\end{equation}
specified by
\begin{equation}
\biggl\{\
z_{w;0}^h\,,\ z_{0;1}^h\,,\ z_{1;1}^h
\
\biggr\}\,.
\eqlabel{susncir}
\end{equation}
Note that in total we have $3+3=6$ parameters (\eqref{susnc} and \eqref{susncir}),
as necessary to specify a solution of a coupled system of 3 second-order ODEs \eqref{on1}-\eqref{on3}.

If the constraint \eqref{wconst} is imposed, it determines $\delta\cala_t'$ algebraically
from $\delta\calv_t'$ and
$\delta W$, and sets 
\begin{equation}
z_{0;2}=0\,,\qquad z_{0;1}^h=-\frac{2(\ (f_{1,0}^h)^4(f_{2,0}^h)^4\ z_{1;1}^h-z_{1;1}^h+12 a_1^h\ z_{w;0}^h\ )}{(f_{1,0}^h)^4(f_{2,0}^h)^4+2}\,.
\eqlabel{z0eli}
\end{equation}

\begin{figure}[ht]
\begin{center}
\psfrag{m}[tt][][1.0][0]{{$\mu$}}
\psfrag{n}[bb][][1.0][0]{{$1/z_{w;2}$}}
\psfrag{c}[tt][][1.0][0]{{$1/z_{w;2}$}}
\includegraphics[width=3in]{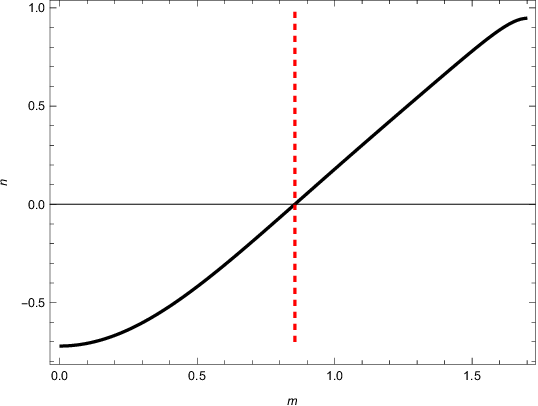}
\includegraphics[width=3in]{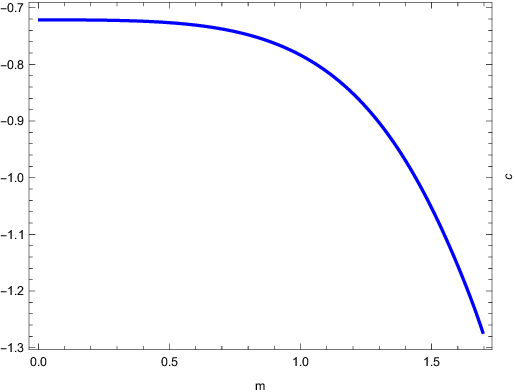}
\end{center}
  \caption{Left panel: the divergence of the normalizable coefficient $z_{w;2}$ (as we vary
  the chemical potential $\mu$ of the baryonic black branes) at $\hat{\mu}_*$ \eqref{hmus}
  (the vertical red dashed line) signals the onset of a new homogeneous phase
  of the charged conifold gauge theory plasma. Right panel: such divergence is absent
  if the additional constraint \eqref{z0eli} is used.
} \label{figure3}
\end{figure}

In fig.\ref{figure3} we present numerical results for the susceptibility $\frac{1}{z_{w;2}}$
as a function of the baryonic chemical potential $\mu$ of the conifold plasma. 
In the setting {\bf (O-I)} (the left panel)  there is a divergence at
\begin{equation}
\hat{\mu}_*=0.855016(8)\,,
\eqlabel{hmus}
\end{equation}
indicating the onset of instability towards a homogeneous plasma phase with
a finite $R$-symmetry charge density (since the value of $z_{0;2}\ne 0$).
Within numerical error, this is precisely the onset of the clumping instability identified
in section \ref{hydro}. Indeed, from \eqref{mus},
\begin{equation}
\frac{\hat{\mu}_*}{\mu_*}-1=5.(3)\times 10^{-7}\,.
\end{equation}
Such an instability can not arise from the hydrodynamic instability, as the susceptibility
computed in the setting {\bf (O-II)} (the right panel) remains finite.

\subsection{$R$-symmetry charged baryonic black branes with  $\mu_B\ne 0$ and  $\mu_R=0$ }\label{normal}

In this section we discuss the construction and the thermodynamics of  black branes of the effective action \eqref{skw1}
with a finite baryonic chemical potential $\mu_B\ne 0$, and a vanishing chemical potential for the $R$-symmetry charge, $\mu_R=0$.
These black branes would, nonetheless, carry the $U(1)_R$ symmetry charge --- they represent the new homogeneous and
isotropic phase of the conifold gauge theory plasma that connects to the HKPT phase at the onset of the instability
discussed in section \ref{onset}.

The corresponding gravitational dual is constructed within 
 the following ansatz:
\begin{equation}
\begin{split}
&ds_5^2=g_{\mu\nu}dx^\mu dx^\nu= -\hat{c}_1^2\ dt^2+\hat{c}_2^2\ d\bm{x}^2+\hat{c}_3^2\ dr^2\,,\qquad \hat{c}_i
=\hat{c}_i(r)\,,\\
&a_1^\Phi=\Phi(r)\ dt\,,\qquad \cala=A(r)\ dt\,,\\
&\calv=V(r)\ dt+\calv_r(r)\ dr\,,\qquad  \{u,v,w\}=\{u,v,w\}(r)\,,
\end{split}
\eqlabel{obackw1}
\end{equation}
where we used the gauge transformations \eqref{gt} to set $a_{1,r}^\Phi=\cala_r=a\equiv 0$.
From the equation of motion 
\begin{equation}
\frac{\delta S_5}{\delta \calv_r}=0\qquad \Longrightarrow\qquad \calv_r\equiv 0\,.
\eqlabel{gfv}
\end{equation}

It is convenient to introduce
\begin{equation}
\begin{split}
&v\equiv \ln f_1\,,\qquad u\equiv \ln f_2\,, \qquad w\equiv \ln H\,,\\
&\hat{c_1}\equiv c_1\ f_1^{1/3}f_2^{4/3}\,,\qquad \hat{c_2}\equiv c_2\ f_1^{1/3}f_2^{4/3}\,,\qquad
\hat{c_3}\equiv c_3\ f_1^{1/3}f_2^{4/3}\,,\\
&c_1=\frac{\sqrt f}{\sqrt r}\,,\qquad c_2=\frac {1}{\sqrt r}\,,\qquad c_3=\frac{s}{2r\sqrt r}\,.
\end{split}
\eqlabel{obackw3}
\end{equation}

The equations of motion for the fields in \eqref{obackw3} are too long to be presented here:
in summary, we have a coupled system of 6 second-order equations for $\{f_1,f_2,\Phi,A,V,H\}$,
and a pair of the first-order equations for $\{f,s\}$.  
These equations are solved subject to the following
asymptotics:
\nxt in the UV, \ie as $r\to 0_+$,
\begin{equation}
\begin{split}
&f=1+ f_4\ r^2+\left(a_{t,1}^2+\frac23 a_2^2\right)\ r^3-2a_{t,1} h_2 a_2\ r^4 +\calo(r^5)\,,
\end{split}
\eqlabel{0uvgen1}
\end{equation}
\begin{equation}
\begin{split}
&s=1-\frac 12 h_2^2\ r^2+\frac{7}{30} a_2^2\ r^3+\left(-95h_2^4+19f_4h_2^2-5a_2a_{t,1}h_2-15 f_{1,8}\right)\ r^4 +\calo(r^5\ln r)\,,
\end{split}
\eqlabel{0uvgen2}
\end{equation}
\begin{equation}
\begin{split}
&\Phi=\mu+a_2\ r-2a_{t,1}h_2\ r^2-a_2h_2^2\ r^3+\biggl(-\frac12 h_2^3 a_{t,1}+\frac14 a_{t,1}h_2f_4+\frac14 a_2 (a_{t,1})^2+\frac32 h_2 a_{t,3}
\\&+\frac{5}{96} a_2^3+\frac14 a_2 f_{1,6}+\frac{1}{40} a_2^3\ \ln r
\biggr)\ r^4
+\calo(r^5\ln r)\,,
\end{split}
\eqlabel{0uvgen3}
\end{equation}
\begin{equation}
\begin{split}
&f_1=1-\frac 52h_2^2\ r^2+\left(f_{1,6}+\frac{1}{10} a_2^2\ \ln r\right)\ r^3+f_{1,8}\ r^4+\calo(r^5\ln r)\,,
\end{split}
\eqlabel{0uvgen5}
\end{equation}
\begin{equation}
\begin{split}
&f_2=1+\frac12h_2^2\ r^2+\left(
-\frac14 f_{1,6}-\frac{1}{40} a_2^2-\frac{1}{40} a_2^2\ \ln r\right)\ r^3+ \biggl(\frac 12 a_{t,1}a_2h_2-\frac 32 h_2^2f_4+9h_2^4\\
&+f_{1,8}\biggr)\ r^4+\calo(r^5\ln r)\,,
\end{split}
\eqlabel{0uvgen6}
\end{equation}
\begin{equation}
\begin{split}
&A=3  a_{t,1}\ r-4 h_2 a_2\ r^2+5 h_2^2 a_{t,1}\ r^3+\left(4 h_2^3 a_2+\frac12 a_2 h_2 f_4+\frac{7}{10} a_2^2 a_{t,1}\right) r^4+\calo(r^5\ln r)\,,
\end{split}
\eqlabel{0uvgen7}
\end{equation}
\begin{equation}
\begin{split}
&V=-h_2 a_2\ r^2+\left(\frac32 a_{t,3}-\frac52 h_2^2 a_{t,1}\right)\ r^3
+\biggl(-14 h_2^3 a_2+\frac12 a_2 h_2 f_4-\frac{13}{40} a_2^2 a_{t,1}-\frac32 a_{t,1} f_{1,6}\\&-\frac{3}{20} a_{t,1} a_2^2\ \ln r\biggr) r^4+\calo(r^5\ln r)\,,
\end{split}
\eqlabel{0uvgen8}
\end{equation}
\begin{equation}
\begin{split}
&H=1+h_2\ +\frac12 h_2^2\ r^2+ \left(-\frac32 h_2^3-\frac14 h_2 f_4-\frac14 a_2 a_{t,1}\right)\ r^3
+\biggl(-\frac{13}{8} h_2^4-\frac14 h_2^2 f_4\\&-\frac{7}{60} a_2^2 h_2-\frac14 a_{t,1} h_2 a_2+\frac12 h_2 f_{1,6}+\frac{1}{20} h_2 a_2^2\ \ln r\biggr)\ r^4
+\calo(r^5\ln r)\,,
\end{split}
\eqlabel{0uvgen9}
\end{equation}
specified by 
\begin{equation}
\biggl\{\
a_2\,,\, f_4\,,\, f_{1,6}\,,\,  f_{1,8}\,,\ a_{t,1}\,,\ a_{t,3}\,,\ h_2
\
\biggr\}\,,
\eqlabel{0uvpar}
\end{equation}
as functions of a $U(1)_B$ chemical potential $\mu$; 
\nxt in the IR, \ie as $y\equiv 1-r\to 0_+$,
\begin{equation}
\begin{split}
&f_1=f_{1,0}^h+\calo(y)\,,\qquad f_2=f_{2,0}^h+\calo(y)\,,\qquad  s=s^h_0+\calo(y)\,,\qquad H=h_0^h+\calo(y)\,,\\
&
\Phi=a_1^h\ y+\calo(y^2)\,,\qquad A=(a_{t,1}^h-2a_{j,1}^h)y+\calo(y^2) \,,\qquad V=(a_{t,1}^h+a_{j,1}^h)y+\calo(y^2)\,,
\\
&f=\biggl(\frac{2 (s^h_0)^2}{(f_{2,0}^h)^8 (f_{1,0}^h)^2}-\frac{(a^h_1-a_{j,1}^h)^2 (h^h_0)^4}{2(f_{2,0}^h)^4 (f_{1,0}^h)^2}
-\frac{(a^h_1+a_{j,1}^h)^2}{2(f_{2,0}^h)^4 (f_{1,0}^h)^2 (h^h_0)^4}\biggr)\ y+\calo(y^2)\,,
\end{split}
\eqlabel{0irass}
\end{equation}
specified by 
\begin{equation}
\biggl\{\
s^h_0\,,\, f_{1,0}^h\,,\, f^h_{2,0}\,,\, a_1^h\,,\ a_{t,1}^h\,,\ a_{j,1}^h\,,\ h_0^h
\
\biggr\}\,,
\eqlabel{0irpar}
\end{equation}
again, as functions of a $U(1)_B$ chemical potential $\mu$. 
Note that in total, we have 7 parameters in the UV \eqref{0uvpar}
and 7 parameters in the IR \eqref{0irpar}, precisely as needed to
specify a solution of a coupled system of equations for the background fields: $2\times 6+1\times 2=14$.

Once the regular black brane solutions are constructed, we can use the holographic
renormalization\footnote{In this phase one should be careful with the holographic renormalization: even thought there is
no source term for the bulk mode $H$, the counterterms for the renormalization of the $\Delta=2$ mode are needed
to insure the correct thermodynamics; see e.g., \cite{Buchel:2012gw}.}
to extract their thermodynamic properties:
\begin{equation}
\begin{split}
&2\pi T=\frac{2 s^h_0}{(f_{2,0}^h)^8 (f_{1,0}^h)^2}-\frac{(a^h_1-a_{j,1}^h)^2 (h^h_0)^4}{2(f_{2,0}^h)^4 (f_{1,0}^h)^2s^h_0}
-\frac{(a^h_1+a_{j,1}^h)^2}{2(f_{2,0}^h)^4 (f_{1,0}^h)^2 (h^h_0)^4s^h_0}\,,\quad
\hat\cale\equiv 2\kappa_5^2 \cale =-3 f_4\,,\\
&\hat\rho_B\equiv  2\kappa_5^2 \rho_B =-4 a_2\,,\qquad 
\hat{\cals}\equiv 2\kappa_5^2 \cals =4\pi f^h_{1,0} (f^h_{2,0})^4\,,\quad
\hat\Omega\equiv 2\kappa_5^2 \Omega =4 \mu a_2 -3 f_4 -T\hat\cals\,,\\
&\hat\rho_R\equiv  2\kappa_5^2 \rho_R =-2 a_{t,1}\,,\qquad \hat\calo_2\equiv 2\kappa_5^2\calo_2=h_2\,,
\end{split}
\eqlabel{othermo}
\end{equation}
where $T$ is the temperature,  $\cals$ is the entropy density, $\Omega$ is the Gibbs free energy density,
$\cale$ is the energy density, $\rho_B$ is the $U(1)_B$ symmetry charge density,
$\rho_R$ is the $U(1)_R$ symmetry charge density (it is nonzero even though the corresponding
chemical potential $\mu_R=0$), and $\calo_2$ is the thermal expectation value of the  
dimension-2 operator dual to the bulk scalar $H$.
The thermodynamics of R-charged baryonic black branes is explored in section \ref{phase}.

\begin{figure}[ht]
\begin{center}
\psfrag{t}[tt][][1.0][0]{{$T/\mu_B$}}
\psfrag{a}[bb][][1.0][0]{{eq.\eqref{thermconst}$-(A)$}}
\psfrag{b}[tt][][1.0][0]{{eq.\eqref{thermconst}$-(B)$}}
\includegraphics[width=3in]{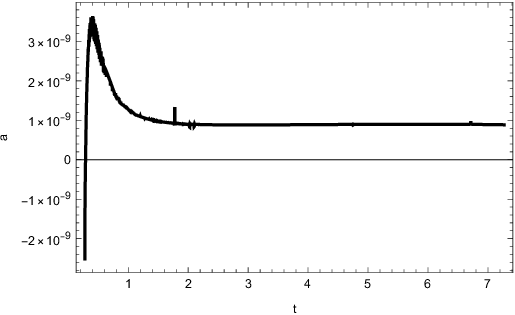}
\includegraphics[width=3in]{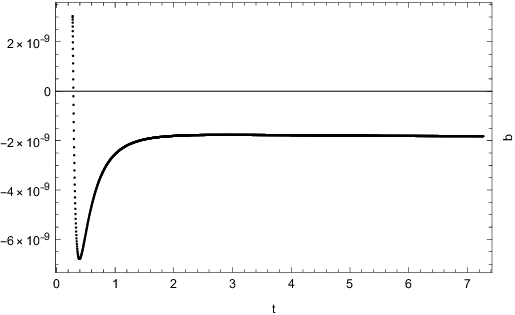}
\end{center}
  \caption{Numerical construction of $R$-charged baryonic black branes is in excellent
  agreement with the expected thermodynamic constraints of eq.\eqref{thermconst}.
} \label{figure4}
\end{figure}

An important check on the numerics is the verification of the thermodynamic constraints
\eqref{thermconst}, presented in fig.~\ref{figure4}. The left/right panel verifies constraint $(A)/(B)$
of \eqref{thermconst}.

\subsection{Non-perturbative instability of the baryonic black branes}\label{nonpert}

It is known that the baryonic black branes reviewed in section \ref{hkpts} suffer from the
nonperturbative D3-$\overline{\rm D3}$ nucleation instability, provided
\cite{Herzog:2009gd}\footnote{See also \cite{Henriksson:2019ifu,Henriksson:2024hsm}
for a related later work.}
\begin{equation}
T\lesssim 0.2\sqrt {2}\ \mu\,.
\eqlabel{approxins}
\end{equation}
The approximate sign in \eqref{approxins} does not allow to establish whether
the critical temperature for the onset of this nucleation instability is above/below
the critical temperature for the $U(1)_R$ symmetry charge clumping instability \eqref{mus}.
In this section we provide the precise evaluation of the critical temperature \eqref{approxins}.

The effective action for the probe $D3/\overline{D3}$ brane in generic type IIB
supergravity background was studied in \cite{Buchel:2004rr}.
For a static\footnote{To compute a 3-brane potential
it is sufficient to consider a 3-brane at a fixed radial location.} 3-brane of charge $q$ (we use $q=+1$ for a $D3$ brane and
$q=-1$ for an $\overline{D3}$ brane) with a world-volume along $\del\calm_5$:
\begin{equation}
S_{3-brane}=-T_3 \int_{\del\calm_5} d^4 x \sqrt{\hat g}+q T_3 \int_{\del\calm_5} \hat{C}_4\,,
\eqlabel{s3}
\end{equation}
where $T_3$ is the 3-brane tension; $\hat g$ is the induced
metric on the world-volume of the probe, see \eqref{n10d} 
with $ds_5^2$ given by \eqref{backw1}, \eqref{backw3},
\begin{equation}
\hat g_{\mu\nu}={\rm diag}(-c_1^2\,,\, c_2^2\,,\, c_2^2\,,\, c_2^2)={\rm diag}
\left(-\frac fr\,,\,\frac 1r\,,\,\frac1r\,,\,\frac1r\right)\,,
\eqlabel{defgh}
\end{equation}
and
\begin{equation}
\begin{split}
d\hat{C}_4=-\frac{2s}{f_1f_2^4 r^3}\ {\rm vol}_{R^{3,1}}\wedge dr\,.
\end{split}
\eqlabel{c4}
\end{equation}
From \eqref{s3}, the 3-brane effective potential $\hat{V}_q\equiv \frac{V_q}{T_3}$ is thus
\begin{equation}
\hat{V}_q=\frac{\sqrt{f}}{r^2}+q \int^r d\rho\ \frac{2s}{f_1f_2^4\rho^3} +{\rm const}\,,
\eqlabel{defepot}
\end{equation}
where the additive constant is such that the $D3$ brane potential $\hat{V}_+$ vanishes at the
AdS boundary, \ie as $r\to 0$. It is easy to see that the $\overline{D3}$ brane potential
is always minimized at the baryonic black brane horizon; hence, a nucleated at some fixed
radial location $\overline{D3}$ brane will move towards the horizon.
In what follows we study the $D3$ brane effective potential $\hat{V}_+$.

\begin{figure}[ht]
\begin{center}
\psfrag{x}[tt][][1.0][0]{{$T/\mu_B$}}
\psfrag{a}[tt][][1.0][0]{{$r$}}
\psfrag{y}[bb][][1.0][0]{{$\hat{V}^h_+$}}
\psfrag{b}[tt][][1.0][0]{{$\hat{V}_+$}}
\includegraphics[width=3in]{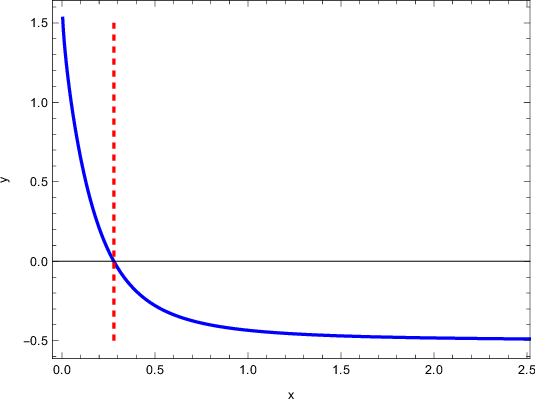}
\includegraphics[width=3in]{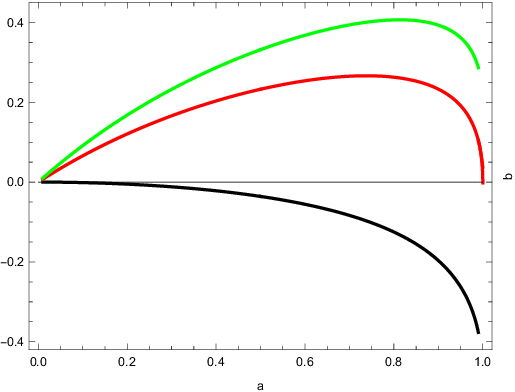}
\end{center}
  \caption{Potential of the probe $D3$ brane at the baryonic black branes
  horizon as a function of $T/\mu_B$ (the left panel). The vertical red dashed line
  is the critical temperature for the onset of the D3-$\rm {\overline{D3}}$ nucleating instability,
 \eqref{vzero}. The right panel: the typical  $D3$ brane potential $\hat{V}_+(r)$ for $\hat{V}^h_+<0$
 (the black curve), for $\hat{V}^h_+=0$
 (the red curve),  and for $\hat{V}^h_+>0$
 (the green curve). The radial coordinate $r$ ranges from the asymptotic AdS boundary $r\to 0_+$,
 to the baryonic black brane horizon $r\to 1_-$.
} \label{figure3a}
\end{figure}

In practice, given a black brane background \eqref{backw1}-\eqref{irpar},
we numerically solve the differential equation (from \eqref{defepot}),
\begin{equation}
0=\frac{d\hat{V}_+}{dr}-\frac{f'}{2 r^2 \sqrt{f}}+\frac{2\sqrt{f}}{r^3}-\frac{2s}{f_1f_2^4r^3}\,,
\eqlabel{dv}
\end{equation}
subject to the following asymptotics:
\nxt in the UV, \ie as $r\to 0_+$,
\begin{equation}
\hat{V}_+=a_2^2\ r +\left(-\frac 18 f_4^2-20 f_{1,8}\right)\ r^2+\calo(r^3)\,;
\eqlabel{potuv}
\end{equation}
\nxt in the IR, \ie as $y\equiv 1-r\to 0_+$,
\begin{equation}
\hat{V}_+=\hat{V}^h_++\frac{\sqrt{2(s_0^h)^2-(a_1^h)^2 (f_{2,0}^h)^4}}{f_{1,0}^h(f_{2,0}^h)^4}\ \sqrt{y}+\calo(y)\,,
\eqlabel{potir}
\end{equation}
where $\hat{V}^h_+$ is the value of the probe $D3$-brane potential at the horizon of the baryonic
black branes. The value of $\hat{V}_+^h$ as a function of the baryonic chemical
potential $\mu$ is presented in fig.~\ref{figure3a} (the left panel). A shape of a
probe $D3$ brane potential crucially
depends on whether $\hat{V}^h_+<0$ or $\hat{V}^h_+>0$, fig.~\ref{figure3a} the right panel.
In the former case,
the global minimum of the potential is at the horizon, and 
it is near the horizon where the  D3-$\rm {\overline{D3}}$ probe brane pair will predominantly
be nucleated, with both branes 'falling through the horizon' --- in this case the
baryonic black brane horizon is free from the ``Fermi seasickness'' of \cite{Hartnoll:2009ns}.
On the contrary, when $\hat{V}^h_+>0$, the global minimum of the $D3$ brane potential
is at the AdS boundary; as a result, for a  D3-$\rm {\overline{D3}}$ probe brane pair nucleation
near the AdS boundary, the $D3$ brane would escape to the asymptotic boundary, while the $\overline{D3}$
will move towards the horizon, leading to the ``Fermi seasickness''. The onset of this instability
is precisely the value of the chemical potential, such that
\begin{equation}
\hat{V}_+^h\bigg|_{T/\mu}=0\qquad \Longleftrightarrow\qquad \frac{T}{\mu}
=\frac{T}{\mu_B}\bigg|_{non-pert}=0.2789(9)\,,
\eqlabel{vzero}
\end{equation}
represented by the vertical red dashed line on the left panel of fig.~\ref{figure3a}.

\subsection{Phases of the conifold gauge theory with $\mu_B\ne0$ and $\mu_R=0$}\label{phase}

Strongly coupled conformal Klebanov-Witten gauge theory at finite baryonic chemical potential and vanishing $R$-symmetry chemical potential
has two phases: a {\it disordered} phase (originally constructed in \cite{Herzog:2009gd} and reviewed in section \ref{hkpts} ),
and the {\it ordered} phase constructed in section \ref{normal}. The ordered phase is called so, as it has a non-vanishing
thermal expectation value of the dimension-2 operator; furthermore, it has a nonzero $R$-symmetry charge.

\begin{figure}[ht]
\begin{center}
\psfrag{t}[tt][][1.0][0]{{$T/\mu_B$}}
\psfrag{w}[bb][][1.0][0]{{$\hat\Omega/\mu_B^4$}}
\psfrag{s}[tt][][1.0][0]{{$\hat\cals/\hat\rho_B$}}
\psfrag{e}[tt][][1.0][0]{{$\hat\cale/(\hat\rho_B)^{4/3}$}}
\includegraphics[width=3in]{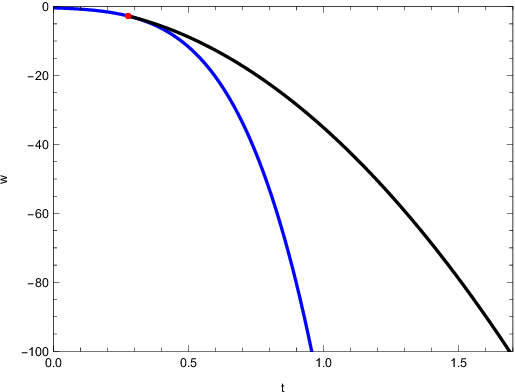}
\includegraphics[width=3in]{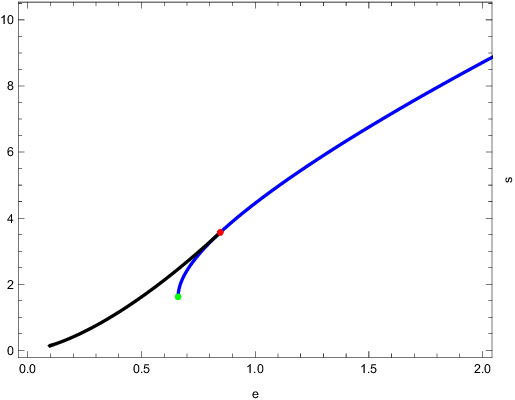}
\end{center}
  \caption{The Gibbs free energy densities $\hat\Omega$ (the right panel) and the entropy densities
  $\hat\cals$ of the disordered (the solid blues curves) and the ordered (the solid black curves )
  phases as  functions of temperature $T$ and the energy densities $\hat\cale$ correspondingly.
  The baryonic chemical potential $\mu_B$ (or the baryonic charge density $\hat\rho_B$) is kept fixed.
The red dot represents the critical temperature/energy density \eqref{tmuc}. The green dot (the right panel)
represents the extremal limit of the baryonic black branes.
} \label{figure1}
\end{figure}

\begin{figure}[ht]
\begin{center}
\psfrag{t}[tt][][1.0][0]{{$\ln[(T-T_{crit})/\mu_B]$}}
\psfrag{o}[tt][][1.0][0]{{$\ln[\hat\calo_2/\mu_B^2]$}}
\psfrag{w}[bb][][1.0][0]{{$\ln[(\hat\Omega_{ord}-\hat\Omega_{dis})/\mu_B^4]$}}
\includegraphics[width=3in]{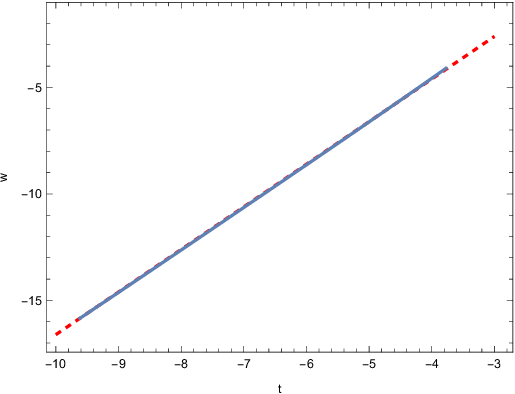}
\includegraphics[width=3in]{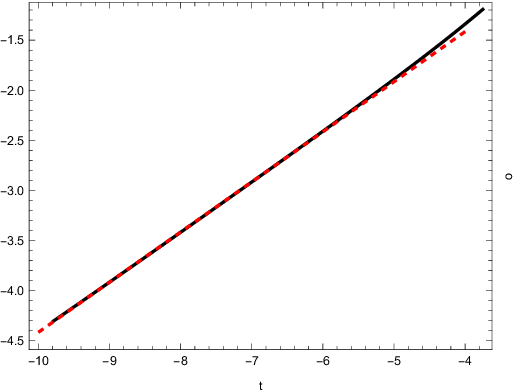}
\end{center}
  \caption{The near-critical behavior, $(T-T_{cirt})\ll T_{crit}$ of the disordered and ordered phases (the left panel).
  The right panel shows the critical behavior of the order parameter: the expectation value of the dimension-2 operator $\hat\calo_2$.
  The red dashed lines have slopes implying the scaling relation as in \eqref{oddiff}.
} \label{figure5}
\end{figure}

\begin{figure}[ht]
\begin{center}
\psfrag{t}[tt][][1.0][0]{{$\ln[T/\mu_B]$}}
\psfrag{o}[tt][][1.0][0]{{$\ln[\hat\calo_2/T^2]$}}
\psfrag{w}[bb][][1.0][0]{{$\ln[\hat\cale/\mu_B^4]$}}
\includegraphics[width=3in]{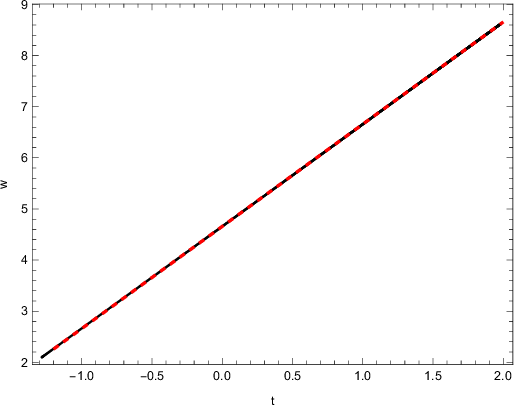}
\includegraphics[width=3in]{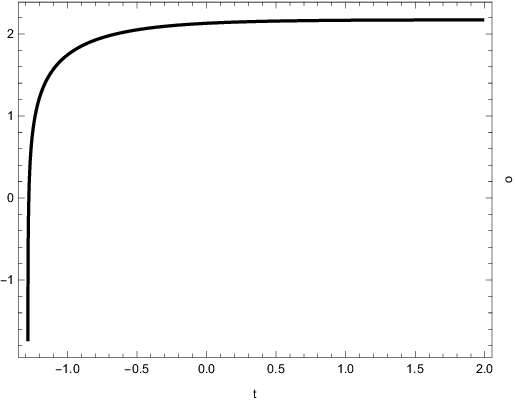}
\end{center}
  \caption{The scaling of the ordered phase energy density $\hat\cale$ (the left panel) and the order parameter $\hat\calo_2$
  (the right panel) as $T\gg \mu_B$. The slope of the red dashed line implies the energy density scaling \eqref{scl1}.
} \label{figure6}
\end{figure}

\begin{figure}[ht]
\begin{center}
\psfrag{t}[tt][][1.0][0]{{$T/\mu_B$}}
\psfrag{l}[tt][][1.0][0]{{$\ln[(T-T_{crit})/\mu_B]$}}
\psfrag{r}[tt][][1.0][0]{{$\hat\rho_R/\hat\rho_B$}}
\psfrag{u}[bb][][1.0][0]{{$\ln[\hat\rho_R/\hat\rho_B]$}}
\includegraphics[width=3in]{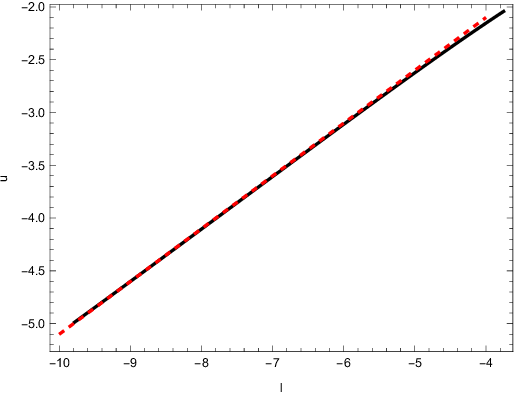}
\includegraphics[width=3in]{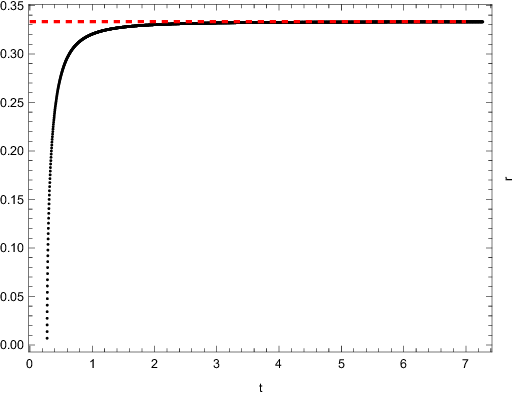}
\end{center}
  \caption{$R$-symmetry charge density $\hat\rho_R$ in the ordered phase is yet another
 order parameter distinguishing this phase from the disordered one (the left panel).
 The slope of the red dashed line implies the scaling relation \eqref{scl4}. 
  As $T\gg \mu_B$ the baryonic and the $R$-symmetry charge densities become proportional
  to each other (the right panel). The horizontal dashed red line is at $\frac 13$.
} \label{figure7}
\end{figure}

The phase diagram of the conifold gauge theory at finite baryonic chemical potential $\mu_B$ is presented in fig.~\ref{figure1}.
The red dot indicates the critical temperature \eqref{tmuc}. The disordered phase (the solid blue lines)
exists for arbitrary temperatures;
in the extremal limit $\frac{T}{\mu_B}\to 0$ is has finite energy and entropy densities (the right panel, green dot).
The ordered phase (the solid black lines) is exotic \cite{Buchel:2009ge,Buchel:2017map,Buchel:2018bzp}
--- it originates at the criticality, however it extends to higher, rather than the lower temperatures.
The ordered phase is subdominant in the grand canonical ensemble (the left panel); however,
it has higher entropy density at the same value of the baryonic charge density than the disordered phase (the right panel).
The right panel is not a genuine microcanonical ensemble since the disordered phase always
has zero charge density of the $U(1)_R$ symmetry, while the ordered phase carries a nontrivial charge density
$\hat\rho_R\ne 0$. 

Fig.~\ref{figure5} explores the critical regime: the left panel shows the difference
between the Gibbs free energy densities of the ordered and the disordered phases, and the right panel
shows the thermal expectation value of the dimension-2 operator $\hat\calo_2$ in the near-critical regime
of the ordered phase. The red dashed lines have slopes $2$ and $\frac 12$ in the left/right panels
correspondingly. Thus, close to criticality, \ie for $(T-T_{crit})\ll T_{crit}$,
\begin{equation}
(\hat\Omega_{ord}-\hat\Omega_{dis})\propto \mu_B^2 (T-T_{crit})^2\,,\qquad \hat\calo_2\propto \mu_B^{3/2} \sqrt{T-T_{crit}}\,.
\eqlabel{oddiff}
\end{equation}

The ordered phase appears to extend to arbitrary high temperatures - it is another example of the conformal ordered
phase in the presence of chemical potential \cite{Buchel:2025cve,Buchel:2025tjq}\footnote{Charge neutral
conformal order was recently studied in \cite{Chai:2020zgq,Buchel:2020xdk,Buchel:2020jfs,Buchel:2020thm,Chai:2021tpt,Chaudhuri:2021dsq,Buchel:2022zxl,Chai:2021djc}.}. In fig.~\ref{figure6} we show the large $\frac{T}{\mu_B}$ scaling of the energy density $\hat\cale$ (the left panel)
and the order parameter $\hat\calo_2$ (the right panel) of the ordered phase. The slope of the red dashed line is 2, implying
that
\begin{equation}
\hat\cale\ \propto \mu_B^2 T^2\,, \qquad {\rm as}\qquad T\gg \mu_B\,,
\eqlabel{scl1}
\end{equation}
while 
\begin{equation}
\hat\calo_2\ \propto T^2\,,\qquad {\rm as}\qquad T\gg \mu_B\,.
\eqlabel{scl2}
\end{equation}
This is identical scaling to that of the ordered phase of the $\caln=4$ SYM plasma \cite{Buchel:2025cve,Buchel:2025tjq}.
We verified that the scaling of the remaining thermodynamic quantities is consistent with the first law of thermodynamics, give \eqref{scl1}:
\begin{equation}
\hat\Omega\ \propto -\mu_B^2 T^2\,,\qquad \hat\cals \propto \mu_B^2 T\,,\qquad \rho_B\propto \mu_B T^2\,.
\eqlabel{scl3}
\end{equation}

The ordered phase has a finite charge density $\hat\rho_R$ under the $U(1)_R$ global symmetry.
This charge density can be viewed as another order parameter distinguishing the disordered and the
ordered phases of the conifold gauge theory.
In the left panel of fig.~\ref{figure7} we show the near-critical scaling of the $R$-symmetry charge density,
while in the right panel we show that  $\rho_R\ \propto \rho_B$ as $T\gg \mu_B$. The slope/offset of the red dashed lines
in the left/right panels imply
\begin{equation}
\frac {\rho_R}{\rho_B} =\begin{cases}
\propto \left(\frac{T-T_{crit}}{\mu_B}\right)^{1/2}\,,\ &{\rm as}\ (T-T_{crit})\ll T_{crit}\,,\\
\frac 13\,,\ &{\rm as}\ T\gg \mu_B\,.
\end{cases}
\eqlabel{scl4}
\end{equation}

\section*{Acknowledgments}
I would like to thank Igor Klebanov for communications, and the referee for encouraging
revising the "Fermi seasickness'' instability of the baryonic black branes. 
Research at Perimeter
Institute is supported by the Government of Canada through Industry
Canada and by the Province of Ontario through the Ministry of
Research \& Innovation. This work was further supported by
NSERC through the Discovery Grants program.

\bibliographystyle{JHEP}
\bibliography{bbb}

\end{document}